\documentclass[12pt]{article}%
\usepackage{amsfonts}
\usepackage{amsthm,amsfonts,amsmath,amssymb,mathtools,graphics}
\usepackage{setspace}

\usepackage{graphicx}
\usepackage{geometry}
\usepackage{hyperref}
\usepackage{footmisc}
\usepackage{enumitem}

\providecommand{\U}[1]{\protect\rule{.1in}{.1in}}
\newcommand{\xnorm}[1]{ \Vert #1 \Vert }
\newcommand{\var}{\text{\rm var}}

\newcommand{\xset}[1]{\left\{ #1 \right\}}

\newtheorem*{berryEsseenTh}{Berry-Esseen
Theorem}
{\bigskip }
\newtheorem*{HoeffdingInequality}{Hoeffding's Inequality}
{\bigskip}
\setlist{nosep}
\newtheorem{theorem}{Theorem}

\newtheorem{assumption}{Assumption}

\newtheorem{corollary}[theorem]{Corollary}

\newtheorem{lemma}[theorem]{Lemma}
\newtheorem{proposition}{Proposition}
\theoremstyle{definition}

\newtheorem{example}{Example}

\DeclareMathOperator*{\argmax}{argmax}
\begin{document}

\title{Efficiency in Collective Decision-Making via Quadratic Transfers\thanks{We
thank Alessandro Lizzeri, Tilman B{\"{o}}rgers, Arunava Sen and audiences at seminars at Princeton U., the U. Michigan, and the 2021 Midwest Theory conference at Michigan State for their helpful suggestions. An earlier version of this work---subsumed into the present paper---circulated under the title "Quadratic voting with multiple alternatives."}}
\author{Jon X. Eguia\thanks{Authors' emails and affiliations: eguia@msu.edu, Michigan
State University (JXE); nicimm@gmail.com, Microsoft (NI);
lalley@galton.uchicago.edu, University of Chicago (SPL);
katrina@cs.huji.ac.il, Hebrew University (KL); glenweyl@microsoft.com,
Microsoft (GW); and xefteris.dimitrios@ucy.ac.cy, University of Cyprus (DX).}
\and Nicole Immorlica
\and Steven P. Lalley
\and Katrina Ligett
\and Glen Weyl
\and Dimitrios Xefteris}
\maketitle

\begin{abstract}
Consider the following collective choice problem: a group of budget
constrained agents must choose one of several alternatives. Is there a budget balanced mechanism that: i) does not depend on the specific
characteristics of the group, ii) does not require unaffordable transfers, and
iii) implements utilitarianism if the agents' preferences are quasilinear and their
private information? We study the following procedure: every agent can express
any intensity of support or opposition to each alternative, by transferring to
the rest of the agents wealth equal to the square of the intensity expressed;
and the outcome is determined by the sums of the expressed intensities. We
prove that as the group grows large, in every equilibrium of this
quadratic-transfers mechanism, each agent's transfer converges to zero, and
the probability that the efficient outcome is chosen converges to one.



\end{abstract}

\newpage\onehalfspacing

\section{Introduction}

Consider a large group of agents who desires to select its collective choice procedure, that is, the mechanism that will be used whenever the group is called to
choose one out of several alternatives. Assume that the group wants this procedure to be efficient---to be able to select the utilitarian outcome related to any collective choice problem it may encounter in the future---in an environment in which each agent has private values and private information about her preferences.
Further, assume that the group wants to make its choices using a mechanism that is budget-balanced; 
that does not require agents to pay large sums (so agents with tight budget constraints can participate); 
and that does not need fine-tuning before each decision (e.g. by taking into account the specific attitudes and beliefs of the group's members regarding the issue at hand).


Classic democratic theory advocates that collective decisions should be made
by voting, endowing each citizen with an equal number of votes (Dahl 1989). Such procedures do not raise issues regarding budget-balance or budget constraints, but to identify the optimal voting rule for any given issue---even in binary contexts---one has to know the relevant attitudes and the beliefs of the group (e.g. Ledyard and Palfrey 2002, Gr\"{u}ner and Tr\"{o}ger 2019). 
Moreover, voting games with multiple alternatives typically feature multiple
equilibria with different outcomes, so the implementation of the utilitarian
alternative is not guaranteed (e.g. Palfrey 1989, Myerson 2002, Bouton 2013, Martinelli and Palfrey 2020). Consequently, to implement
utilitarianism in settings with several candidates, we must consider mechanisms with transfers.

The VCG mechanisms of Vickrey (1961), Clarke (1971) and Groves (1973)
implement utilitarianism in dominant strategies, but are not budget-balanced.
Budget-balance can be attained by running a VCG mechanism with $n-1$ agents,
and letting the $n$-th agent be the residual claimant of all transfers; this
mechanism is asymptotically optimal, in the sense that the probability that it
chooses optimally converges to one in the size of the group (Green and Laffont
1977), but the mechanism requires agents to be able to pay large transfers or
fines, so it cannot run if agents have tight individual budget
constraints.\footnote{Implementing utilitarianism with a budget-balanced
mechanism in complete information settings is known to be feasible if the
implementation concept is Nash equilibrium as in Maskin (1999) or in the
multi-bidding mechanism by Perez-Castrillo and Wettstein (2002); or
undominated Nash equilibrium (Sj\"{o}str\"{o}m 1994); or subgame perfect Nash
equilibrium (Moore and Repullo 1988).}

The expected externality AGV mechanisms by d'Aspremont and Gerard-Varet (1979)
and Arrow (1979) are budget-balanced and implement utilitarianism in
Bayes-Nash equilibrium by charging each agent the expected externality that
her report causes to other agents. This expected externality
vanishes to zero in the size of the group, so asymptotically, the AGV
mechanisms do not require large transfers. 
However, to appropriately charge any given report, the AGV mechanism needs to know the exact externality that the agents expect it to cause.\footnote{The demanding informational environment in which the mechanism knows the agents' common beliefs is known as the payoff type space. The most permissive environment is the universal type space, which is the union of all type spaces that arise from the payoff environment (B\"{o}rgers, 2015). While Bergemann and Morris (2005) make the case for analyzing the latter, they also suggest studying environments like ours: \emph{\textquotedblleft[t]here are many
type spaces in between [...] that
are also interesting to study. For example, we can look at all payoff type spaces
(so that the agents have common knowledge of a prior over payoff types but
the mechanism designer does not)".} Our theory follows this suggestion, as we study collective choices made by agents who have common knowledge of a prior over payoff types, using a mechanism that operates without knowing this prior.} Therefore, if the group adopts this approach to deal with a series of collective choice problems, the procedure would need to be readjusted by a benevolent and informed planner (i.e. someone who has the power to adapt the group's decision-making rule, who cares to implement the utilitarian optimum, and who knows the voters' beliefs) before every decision. In the absence of such a planner, the group might have to resort to a collective choice mechanism that does not demand frequent reconfigurations.\footnote{Notice that if a large collective entity can find a planner with these characteristics, then there is no need to employ voting, VCG, AGV, or any other procedure of preference aggregation: the planner can simply implement the utilitarian optimum by command (Ledyard and Palfrey 2002).}

In this paper we study a quadratic-transfers mechanism (QTM) and a population with quasilinear preferences, and show that QTM asymptotically
implements utilitarianism in Bayes-Nash equilibrium.
Under QTM
agents can express any intensity of support for or opposition to each
alternative, at a cost that is the square of the intensity expressed. These
quadratic costs for the expression of preferences for or against each
alternative are then aggregated across the multiple alternatives to derive the
total costs for the agent. This total cost collected by the mechanism from
each agent is redistributed in the form of a transfer to all other agents. QTM
then chooses the outcome as a function of the total net support for each alternative.

We show that for a sufficiently large population, QTM chooses the utilitarian
optimal outcome with probability arbitrarily close to one, and individual payments converge to zero. That is, since QTM is always budget-balanced and does not use any information about the agents preferences or beliefs, QTM asymptotically satisfies all our stated desiderata.

Apart from the theoretical motivation to study the QTM, there is an additional---and potentially---more pressing reason. Mechanisms with transfers attract increasing media attention, and hence enter into the public debate.\footnote{See \textquotedblleft The mathematical method that could offer a fairer way to vote," in \textit{The Economist} December 18, 2021; \textquotedblleft Digital tools can be a useful bolster to democracy,"in the \textit{Financial Times} February 1, 2020; and "A new way of voting that makes zealotry expensive," in \textit{Bloomberg Businessweek} May 1, 2019.} Indeed, several organizations have already started experimenting with collective choice procedures that redistribute voting power among issues and voters---even in the presence of multiple alternatives---motivated by an efficiency rationale.\footnote{For instance, the Democratic majority caucus in the Colorado House of Representatives used a mechanism with quadratic transfers of voice credits to allocate \$40 million among a number of proposals. See \textquotedblleft \$120 million in requests and \$40 million in the bank. How an obscure theory helped prioritize the Colorado budget," in \textit{The Colorado Sun} May 28, 2019.} Yet, there is little knowledge about what we should expect from such mechanisms in general non-binary settings. Are they likely to deliver higher social welfare compared to standard voting rules? If so, under which conditions and at what cost? Arguably, theoretical studies should anticipate the introduction of novel collective choice procedures into the public discourse, in order to properly inform the stake holders regarding their potential pros and cons. Our work provides relevant insights: we show, for the first time, how a mechanism with quadratic transfers could be designed to admit a unique and efficient equilibrium outcome in a setting with multiple alternatives, and we also identify limitations and caveats associated with its use.
 
QTM draws inspiration from all the mechanisms with transfers that have been proposed in the literature (e.g. the VCG and the AGV), and in particular from those that consider quadratic payments either in ``influence vouchers'' (e.g. Hylland and Zeckhouser 1980) or ``voice credits'' (e.g. Casella and Sanchez 2020), or in private consumption (e.g. Posner and Weyl 2015, Goeree and Zhang 2017, Lalley and Weyl 2018, 2019).\footnote{The interested reader is referred to Casella and Mace (2021), Eguia and Xefteris (2021) and
their literature reviews for additional relevant work in the binary setting.
The suggestion that the votes cast by an entity should be the square root of
its size dates back at least to Penrose (1946). Quadratic costs also arise as
the limit of d'Aspremont and Gerard-Varet's (1979) mechanism as the size of
the group diverges to infinity (Goeree and Zhang 2017).} To our knowledge, though, QTM is the first mechanism with quadratic transfers that implements the utilitarian outcome irrespective of the number of the available alternatives, and that is robust to deviations of a small number of voters to any affordable strategy. The possibility of negative voting is the key in reaching efficiency in settings with multiple candidates: as we argue after the statement of the main result, if voters are not allowed to express opposition to any alternative, then inefficient equilibria may also obtain. Moreover, the outcome function employed by the QTM, assigns a win probability to each alternative that is strictly increasing in its net support. This leads to an arbitrarily large number 
of vote acquisitions in equilibrium as the population grows large,
and hence to an essentially zero effect of each individual voter on the outcome, 
even if one deviates to any affordable action. 
This deals effectively with the issue of a bounded total number of vote acquisitions
encountered in previous equilibrium analyses in binary setups 
with alternative outcome functions (e.g. Lalley and Weyl 2019).\footnote{In binary choice problems, or in elections with two candidates, choosing the winner by majority rule with costly voting is known to lead to high social welfare (B{\"o}rgers 2004) and, under certain conditions, it even leads to asymptotically implementing the utilitarian outcome (Ledyard 1984, Krishna and Morgan 2011, 2015). In non-binary setups, voters may behave truthfully because devising manipulations is computationally infeasible (Conitzer et al. 2007, Faliszewski and Procaccia 2010, Brandt et al. 2016, Meir 2018), and best-response dynamics can lead to outcomes that are nearly optimal according to some scoring rules (Branzei et al. 2013). However, these outcomes obtained by sincere play, or by learning
through best response dynamics, can also fail to attain utilitarian
efficiency.} 

We analyze QTM under the restrictive, but standard, assumption of quasi-linear
preferences, which rules out wealth effects.\footnote{The classic VCG and AGV
mechanisms, the multi-bidding mechanism, and the more recent quadratic
mechanisms mentioned above, together with most of the related literature, also
restrict attention to quasilinearity. We discuss the implications of relaxing
this assumption in Section \ref{sec:disc}.} Therefore, our results are more
relevant to large groups of risk-neutral agents (e.g. the 300,000 firms that are members of the US Chamber of Commerce; shareholders, etc.)
than to groups of risk-averse individuals with unequal wealth, and thus we
refrain from proposing that democracies use QMT as the main mean to aggregate
(risk-averse) citizen's preferences over ideological
choices.\footnote{Storable votes (Casella 2005)\ constitute a possible
alternative for these applications. Notably, a combination of storable votes
with quadratic conversion of stored votes to votes applied to a given decision
has been found to increase welfare in a survey experiment based on four ballot
initiatives in California (Casella and S{\'a}nchez 2022).} Beyond this
restriction, the QTM shares additional limitations with other existing
mechanisms: efficiency need not obtain in small groups, nor if agents'
preferences are correlated, and the mechanism is not immune to collusion by
coalitions of agents.\footnote{Weyl (2017), Lalley and Weyl (2019), and
Goeree, Louis and Zhang (2020) discuss and analyze conditions under which
mechanisms with quadratic transfers do not guarantee efficiency in binary choice.} Hence, our
formal result is best interpreted as basic research that is informative, but
not immediately pertinent, to all settings of applied interest.

\section{Model Assumptions and Main Result}

\label{sec:model-assumpt-main}

A set of $n$ agents faces a collective choice over a finite set of $m \geq 2$ alternatives. Each agent $i$ has a \emph{type} (or \emph{value})
$u^{i}=(u^{i}_{j})_{j=1,2,\cdots,m}\in[0,1]^{m}$, which characterizes her
preferences over (alternative, wealth) pairs. We assume throughout that the
$n$ agents' types are drawn independently from a common probability
distribution $F$ on $[0,1]^{m}$, which is known to all $n$ agents.\footnote{This is a standard assumption when studying elections with private values and information (see, for instance, Palfrey 1989; Ledyard and Palfrey 2002; Myerson 2002; Bouton 2013; or Gr\"{u}ner  and  Tr\"{o}ger 2019).} To emphasize that the agents' values are now to be treated as \emph{random vectors}
we shall denote them by upper-case letters $U^{i}$, and will use lower-case letters
$u^{i}$ to denote particular values of these random vectors.

According the QTM, agents acquire a number of votes, either positive or negative (i.e., for or against), for each
of the $m$ alternatives. To acquire $a_{j}$ votes for (or against) alternative
$j$, an agent must transfer $c a_{j}^{2}$ dollars to the rest of the voters. The unit cost $c>0$ per vote is
a free parameter in the model. Agents may use auxiliary randomization in
determining their vote purchases: thus, each agent $i$'s vote $A^{i}_{j}$ for
each alternative $j \in\{1,...,m\}$ is a random variable defined as a
function
\begin{equation}
\label{eq:vote-rule}A^{i}_{j}:= \alpha^{i}_{j}(U^{i}, Z_{i})
\end{equation}
where each $\alpha^{i}_{j}:[0,1]^{m}\times[0,1]\rightarrow\mathbb{R}$ is a
Borel-measurable function of the agent's type and an auxiliary random variable
$Z_{i}$, where $Z_{1},Z_{2}, \cdots,Z_{n}$ are independent and uniformly
distributed on the unit interval $[0,1]$. (See, e.g, Aumann (1964) for an
explanation of how uniform random variables are used to implement randomized
strategies.) If the functions $\alpha^{i}_{j}$ do not depend explicitly on the
arguments $Z_{i}$, then the agents are said to be using a \emph{pure strategy}.
 
For each agent $i$, let $A^{i}:=(A_{1}^{i},...,A_{m}^i)$ denote agent $i$'s (random) vote vector,
let $A:=(A^1,...,A^2)$ denote the vote profile, 
and let $A^{-i}:=(A^1,...,A^{i-1},A^{i+1},...,A^n)$ denote the voting profile
of all other agents, without $i$.

The agents' payoffs---given their types---are determined by the (random) vote
totals
\begin{equation}
\label{eq:vote-totals}V_{j}:= \sum_{i=1}^{n} A^{i}_{j}%
\end{equation}
and by a further randomization: given the vote totals $V_{j}$, alternative $k$
is chosen with probability
\begin{equation}
\label{eq:e}Q_{k}:=\frac{e^{V_{k}}}{\sum_{j}e^{V_{j}}}.
\end{equation}
The probability distribution $(Q_{j})_{j \in[m]}$ is itself a random variable,
because the vote totals depend on the agents' types $U^{i}$ and the random
variables $Z_{i}$. Agents are expected utility maximizers. The utility for
agent $i$ with type realization $u^{i}$ is
\begin{equation}
\label{eq:u}W^{i} :=\sum_{k} Q_{k}u^{i}_{k} -c\sum_{k}(A^{i}_{k})^{2}+\frac
{c}{n-1}\sum_{i^{\prime}\not =i} \sum_{k}(A^{i^{\prime}}_{k})^{2}.
\end{equation}
Since the type components $U^{i}_{j}$ take values $u^{i}_{j}$ between $0$ and
$1$ (with probability $1$), it is sub-optimal for any agent $i$ to acquire more
than $1/ \sqrt{c}$ votes either for or against any alternative $j$. Therefore,
we shall only consider decision rules for which the domain of the random vote vector $A^{i}$ 
is contained in $\mathcal{A}:=[-c^{-1/2},c^{-1/2}]^{m}$ for each agent $i$.

In addition, we make two assumptions on the sampling distribution $F$.

\begin{assumption}
\label{assumption:F1} The probability distribution $F$ is supported by the
unit cube $[0,1]^{m}$, and its coordinate means
\[
\mu_{j}:=\int u_{j} \, dF(u) ,
\]
where $u_{j}$ denotes the $j$th coordinate of the vector $u$, satisfy $\mu
_{i}\neq\mu_{j}$ for any distinct coordinates $i$ and $j$.
\end{assumption}

Every Pareto optimal outcome must involve choosing alternative $\argmax_{j\in{1,...,m}}\sum^{n}%
_{i=1}U^{i}_{j}$, which maximizes the sum of values. Assumption
~\ref{assumption:F1} \text{ }implies that if $n$ is sufficiently large, this
utilitarian optimum is alternative $\argmax_{j\in{1,...,m}}\mu_{j}$ with
probability arbitrarily close to one.\footnote{Assumption 1 excludes
distributions such that the expected value of each alternative is the same.
Notice, however, that in such symmetric environments welfare concerns
asymptotically vanish, at least per capita. The averaged realized utilities of
each alternative all converge to their common expectation, so in this case any
choice gives asymptotically the same average utility to citizens.

For the case of two alternatives---which is the one that has attracted most attention in the literature---we can show that QTM implements the utilitarian outcome even if voters have incorrect common beliefs regarding the type-generation process (i.e. when types are generated according to $F$, but voters believe that they are generated according to $\hat{F}$). We elaborate on this result, and its extension to multiple alternatives, in Section 4.}

For notational convenience, and without loss of generality, we relabel alternatives, so that their order aligns with the order of their means. That is, we consider
\[
\mu_{1}>\cdots>\mu_{m}.
\]

Note that knowing which alternative is (most likely) the utilitarian one does not resolve society’s collective choice problem. As in Nash implementation environments---where the utilitarian optimum is common knowledge, but implementing utilitarianism is often difficult (Maskin 1999)---, the utilitarian optimum is not directly contractible: no agent can be delegated the task of choosing the utilitarian alternative without risking that this agent will choose instead the alternative she most prefers; rather, if society wishes utilitarian collective choices, then to implement this choice rule it needs a mechanism with the right incentives.

Indeed, important recent contributions have found it worthwhile to seek optimal mechanisms in this exact informational environment, e.g. Gr\"{u}ner and Tr\"{o}ger (2019) in contexts with costly and voluntary participation, and Ledyard and Palfrey (2000) with compulsory voting, or even under a complete information environment, e.g. the multi-bidding mechanism by Perez-Castrillo and Wettstein (2002).

Assumption ~\ref{assumption:F2} \text{ }is a mild technical assumption on the
distribution of types, and is satisfied, for instance by any distribution with
positive density everywhere in a neighborhood of indifference about all alternatives.

\begin{assumption}
\label{assumption:F2} For each coordinate $j=1,2,\cdots,m$ there exists a
\emph{positive} number $t_{j}\in(0,1]$ such that the distribution $F$ assigns
positive probability to every neighborhood of the vector
\[
t_{j}e_{j}=(0,0, \cdots,0,t_{j},0,\cdots,0,0).
\]

\end{assumption}

Our main result describes an asymptotic property of all \emph{symmetric
Bayes-Nash equilibria}, possibly in mixed (randomized) strategies, in
sufficiently large populations. To guarantee that this result is meaningful, we
first establish that at least one such equilibrium exists.

\begin{proposition}
\label{proposition:existence} For every sample size $n$ and every value $c>0$
of the cost-per-vote parameter there is a symmetric Bayes-Nash
equilibrium (possibly in mixed strategies).
\end{proposition}

The proof of this existence---together with the proof of all other intermediate
results---is in the Appendix. Our main result shows that in sufficiently large
collective entities, QTM chooses the utilitarian optimum alternative with probability
converging to one.

\begin{theorem}
\label{theorem:main} Suppose that the sampling distribution $F$ satisfies
Assumptions \ref{assumption:F1}-\ref{assumption:F2}. Then there is a sequence
of real numbers $\beta_{n}\downarrow0$ such that for any sample size $n\geq1$
and any symmetric Bayes-Nash equilibrium,
\begin{equation}
P\left\{  Q_{1}\leq1-\beta_{n}\right\}  \leq\beta_{n}. \label{eq:main}%
\end{equation}

\end{theorem}

In the next section we explain the proof of this result. Here we anticipate an
intuition, as follows.

For each agent $i$ with type $u^i$ and vote vector $a^{i}\in\mathcal{A}$, for each alternative $k$, 
and for a given random voting profile $A^{-i}$, 
define the \textquotedblleft relative valuation\textquotedblright\ of alternative $k$ for agent $i$,
denoted $\omega_{k}^{i}$, as the difference between the value $u_{k}^{i}$ that agent
$i$ obtains if alternative $k$ is chosen, and the value for $i$ if
$k$ is not chosen, given vote vector $a^{i}$ and vote profile $A^{-i}$. Further, let $piv_{k}^{i}$ denote 
the marginal pivotality of vote $a_{k}^{i}$, defined as the derivative of the probability
of choosing $k$ with respect to $a_{k}^{i}$, given $A^{-i}$. 
Agent $i$ chooses vote $a_{k}^{i}$
optimally for each alternative $k$ by equating the marginal cost $2ca_{k}^{i}$
to the expected value of $\omega_{k}^{i}\cdot piv_{k}^{i}$. If
$piv_{k}^{i}$ were a constant across agents, across alternatives and across vote profiles, 
it would follow immediately that optimal votes would be proportional to 
the expected relative valuations,
and thus the alternative with greatest aggregate expected relative valuation would also accrue
the greatest aggregate vote. Alas, the marginal pivotality $piv_{k}^{i}$
varies across alternatives and across vote vectors, introducing two distortions.

The first distortion is that an agent's marginal pivotality varies with the
agent's own vote vector (that is, $piv_{k}^{i}$ varies with $a^{i}$), and thus two
agents with different types (who acquire different quantities of votes), face
different incentives. This first distortion vanishes asymptotically in the
size of the group:\ in a sufficiently large collective entity, the expected total
number of votes acquired by all the players diverges to infinity, 
and the marginal pivotality of each agent becomes arbitrarily close to
zero for almost all realizations of the actions of the other players. So
individual equilibrium votes, for all agents and all types, also converge to
zero. As all individual votes converge, the distortion based on heterogeneity
in vote vectors across agents vanishes:\ optimal vote purchases for all agents, types and
alternatives converge to being proportional to the expected value of
$\omega_{k}^{i}\cdot piv_{k}$, for a pivotality parameter $piv_{k}$ that
varies across alternatives but is common across all agents. This pivotality
parameter $piv_{k}$ captures the relative pivotality of voting for different
alternatives. While all pivotalities converge to zero, $piv_{k}$ marks the
rate at which they converge for each alternative.

The second distortion is that alternatives with the greatest pivotality
attract more votes. Under plurality rule, this leads to the so-called
Duvergerian (1951) dynamics: votes for the expected winner 
or for the expected runner up have the highest pivotality.
Voters perceive that votes for other alternatives are wasted, and thus they cast their vote
either to the front runner or to the main challenger, based on their ordinal preference
among these two. This reasoning can sustain a self-enforcing horse
race between any pair of alternatives, so multiple equilibria exist, 
and none of them is guaranteed to elect the utilitarian
optimum.\footnote{For the exact same reason, multiple Duvergerian equilibria 
also arise under the quadratic voting mechanisms proposed for binary decision-making contexts 
(Goeree and Zhang 2017, Lalley and Weyl 2019), if applied to choices over multiple alternatives. 
If purchased votes can be distributed among the alternatives 
but cannot be cast against any of them, then every voter has incentives 
to cast all her votes to one of the two front runners.} Under QTM, the option to cast
any number of votes (at a linear marginal cost) in favor or against any
alternative dismantles this Duvergerian trap and leads, asymptotically, to the
election of the utilitarian optimum.\footnote{Negative voting under fixed
scoring rules similarly rules out undesirable Duvergerian equilibria (Myerson
2002, section 4), but it sustains insufficiently
discriminatory equilibria in which multiple alternatives are chosen with a
substantial probability (Myerson 2002, section 5).
These insufficiently discriminatory equilibria cannot arise under QTM,
because voters are free to choose the exact number of votes 
that they cast in favor or against any alternative.}

Consider an illustrative example with three alternatives $\{1,2,3\}$ with
expected valuations $\mu_{1}>\mu_{2}>\mu_{3}$.
Under plurality rule, there are three Duvergerian equilibria, each pitting a different pair
of alternatives in a horse race. In contrast, under QTM, if the group is sufficiently
large so that mean valuations converge to their expectation, the aggregate
relative valuation for alternative $3$ in equilibrium must be negative
(because the average of expected values conditional on not choosing alternative
$3$ is at least $\mu_{2}$); whereas, the sum of relative valuations for
alternative $1$ must be positive (because the average of expected values
conditional on not choosing alternative $1$ is at most $\mu_{2}$). Hence, the
equilibrium aggregate vote is positive for alternative $1$ and negative for
alternative $3$, so alternative $1$ wins with greater probability than
alternative $3$. Since the total number of votes diverges as
the population grows large, it follows that the probability that alternative $1$ wins
becomes arbitrarily larger than the probability that alternative $3$ wins.
Therefore, the aggregate relative valuation for alternative $2$ is then also
negative (because the average of expected values conditional on not choosing
alternative $2$ becomes arbitrarily close to $\mu_{1}$ and hence larger than $\mu_{2}$). 
It follows that in every equilibrium, the utilitarian optimum (alternative $1$)
is chosen with the greatest probability.\footnote{As we shall prove below, this probability converges to one
in the size of the population because the vote total for the utilitarian
alternative becomes arbitrarily large if the collective entity is sufficiently large.} 
The intuition in this example generalizes to choice problems with any number
of alternatives, as we demonstrate in the next section. 

\section{Proof of Theorem \ref{theorem:main} \label{sec:proof}}

In this section we outline the structure of the proof, we state all key
intermediate results and we provide partial intuition for them, with a
complete proof of each intermediate step in the Appendix.

After detailing terminology and notation, our proof proceeds in six steps: in
Step 1, we derive the first order conditions on the agents' optimization
problem; in steps two and three, we show that marginal pivotalities vanish
(Step 2), and that as they vanish, their relative values across types converge
(Step 3); in Step 4 we derive approximate bounds on equilibrium behavior, and
in Step 5 we use them to show that pivotalities vanish slowly; finally in Step
6 we use these intermediate results to conclude that alternative $1$ (which is
asymptotically almost always the utilitarian optimum) is almost always chosen
in sufficiently large groups. We derive this last step separately for the
special cases of $m=2$ and $m=3$ alternatives, and then for the more general
case with $m\geq3$ alternatives.

\noindent\textbf{Terminology and Notation.} \newline Since only symmetric
rules will be considered, we shall use the abbreviated term \emph{Bayes-Nash
equilibrium}, or \emph{BN equilibrium}, to mean \emph{symmetric
Bayes-Nash equilibrium} in mixed strategies. The crucial feature of a
symmetric strategy in our setting is that the random vectors
\[
(U^{1},A^{1}), (U^{2},A^{2}), \cdots,(U^{n},A^{n})
\]
are independent and identically distributed; we will use this fact repeatedly.
For any agent $i$ and alternative $j$ we will denote by
\begin{equation}
\label{eq:one-out}V^{-i}_{j}:=\sum_{i^{\prime}\,:\,i^{\prime}\not =%
i}A^{i^{\prime}}_{j}%
\end{equation}
the sum of the votes of the agents $i^{\prime}\not =i$. We will also adopt the
following standard mathematical notation: for any real number $x$,
\begin{align*}
x_{+}  &  := \max(x,0),\\
x_{-}  &  := \max(-x,0),
\end{align*}
and for every positive integer $k$,
\[
[k]:= \left\{  1,2, \cdots,k \right\}  .
\]

The selection probabilities $Q_{j}$ depend on the agents' votes; consequently,
since each agent $i$ knows his/her own type $U^{i}=(U^{i}_{j})_{j \in[m]}$ and
vote vector $A^{i}=(A^{i}_{j})_{j\in[m]}$, in maximizing expected utility
agent $i$ must take expectations \emph{conditional} on $U^{i}$ and $A^{i}$. To
facilitate arguments involving such conditional expectations and
probabilities, we will use the notation $E_{i}$ to denote conditional
expectation given $U^{i}$ and $A^{i}$; thus, for any random variable $Y$, we
write $E_{i}Y$ as shorthand for $E(Y\,|\,\mathcal{F}_{i})$, where
$\mathcal{F}_{i}$ is the $\sigma-$algebra generated by $U^{i}$ and $A^{i}$. By
a standard result in measure theory, any conditional expectation given the
$\sigma-$algebra$\mathcal{F}_{i}$ is a Borel-measurable function of
$(U^{i},A^{i})$. When necessary, we will use the notation
\[
\label{eq:condExp} E(Y \,|\, U^{i}=u^{i}, A^{i}=a^{i})
\]
to indicate the dependence on particular values $u^{i},a^{i}$ of these
conditional expectations.

For each agent $i$, each alternative $k$, and every possible vote vector $a
\in\mathcal{A}$, define
\begin{equation}
\label{eq:Qija}Q^{i}_{k}(a):= \frac{e^{a_{k}}e^{V^{-i}_{k}}}{\sum_{j}e^{a_{j}%
}e^{V^{-i}_{j}}};
\end{equation}
this would be the selection probability for alternative $k$ if agent $i$'s
vote were $A^{i}=a$, and so $Q_{k}=Q^{i}_{k}(A^{i})$ for each $i\in[n]$.

\medskip\noindent\textbf{Step 1. First Order Conditions}. \newline Because the
random vectors $(A^{i})_{ i \in[n]}$ are i.i.d., for any $k \in[m]$ and any $a
\in\mathcal{A}$ the random variable $Q^{i}_{k}(a)$ has the same distribution
as $Q^{1}_{k}(a)$; consequently,
\begin{align}
\label{eq:rjka}E_{i}Q_{k}  &  = E_{i}Q^{i}_{k}(A^{i})= q_{k}(A^{i}) \quad &
\text{where} \quad q_{k}(a):= E Q^{1}_{k}(a), \;\; \text{and}\\
E_{i}Q_{j}Q_{k}  &  = E_{i}Q^{i}_{j}(A^{i})Q^{i}_{k}(A^{i})= r_{j,k}(A^{i})
\quad & \text{where} \quad r_{j,k}(a):= E Q^{1}_{j}(a) Q^{1}_{k}(a).
\end{align}
Elementary calculus show that
\begin{equation}
\label{eq:partials}\frac{\partial q_{k}(a)}{\partial a_{j}}= - r_{j,k}(a)
\quad\text{for} \quad j\not =k, \;\; \text{and}\quad\frac{\partial q_{j}%
(a)}{\partial a_{j}}= q_{j}(a)-r_{j,j}(a).
\end{equation}
(Differentiation under the expectation is justified by the fact that the
derivative of the integrand is uniformly bounded for $a^{i}\in[-1/\sqrt
{c},1/\sqrt{c}]^{m}$; see, for instance, Royden (1988), sec. 4.4, problem 19.)

In a Bayes-Nash equilibrium, agent $i$'s strategy is concentrated on
\emph{best response} vectors $a^{i}\in\mathcal{A}$, that is, vectors that
maximize conditional expected utility $E_{i}W^{i}$ given his/her type and
action $(U^{i},A^{i})=(u^{i},a^{i})$. This conditional expected utility is
given by
\[
E(W^{i}\,|\, (U^{i},A^{i})=(u^{i},a^{i}))= \sum_{k} EQ^{i}_{k}(a^{i})u^{i}_{k}
-c\sum_{k}(a^{i}_{k})^{2}+\frac{c}{ n-1}\sum_{i^{\prime}\not =i} \sum
_{k}E(A^{i^{\prime}}_{k})^{2}.
\]
First-order conditions are gotten by setting partial derivatives of these
expected utilities with respect to the variables $a^{i}_{j}$ to zero. Using
equations \eqref{eq:partials}, we conclude that
\begin{equation}
\label{eq:foc-A}2ca^{i}_{j}= \sum_{k} (u^{i}_{j}-u^{i}_{k}) r_{j,k}(a^{i})
\end{equation}
is a necessary condition for a best response. Observe that the term $k=j$ in
this sum vanishes; thus, $\sum_{k}$ can be replaced by $\sum_{k\,:\, k \not =
j}$. Since in a BN equilibrium an agent $i$'s action $A^{i}$ is with
probability one a best response, we deduce from \eqref{eq:foc-A} and
\eqref{eq:rjka} that with probability $1$,
\begin{equation}
\label{eq:foc}\boxed{ 2cA^{i}_{j}= \sum _{k} (U^{i}_{j}-U^{i}_{k}) E_{i}(Q_{j}Q_{k}).}
\end{equation}

\medskip\noindent\textbf{Step 2. Marginal pivotalities vanish}. \newline The
central mathematical difficulty in characterizing Bayes-Nash equilibria is
that agents with different type realizations have different conditional
expectations $E_{i}(Q_{j}Q_{k})$. Let's call these the \textbf{marginal
pivotalities}. In this subsection we show that as the sample size
$n\rightarrow\infty$ the marginal pivotalities become arbitrarily small.
First, Lemma \ref{lemma:marg-piv} establishes bounds for the ratio of the
conditional and unconditional expectations of pivotality.

\begin{lemma}
\label{lemma:marg-piv} In any BN equilibrium, for every pair $j,k$ of distinct
alternatives, with probability $1$,
\begin{equation}
\label{eq:marg-piv}e^{-16/\sqrt{c} } \leq\frac{E_{i}(Q_{j}Q_{k})}{E(Q_{j}%
Q_{k})}\leq e^{16/ \sqrt{c}}.
\end{equation}
Equivalently, for any vote vector $a\in\mathcal{A}$ and any pair $j,k$ of
alternatives,
\begin{equation}
\label{eq:q-marg-piv}e^{-16/\sqrt{c} } \leq\frac{r_{j,k}(a)}{E(Q_{j}Q_{k}%
)}\leq e^{16/ \sqrt{c}}.
\end{equation}

\end{lemma}

The unconditional expectations $E(Q_{j}Q_{k})$ are determined by the random
vector of vote totals $(V_{1},V_{2},...,V_{n})$. Each vote total $V_{j}$ is a
random variable constructed as the sum of $n$ independent individual actions
$(A^{1}_{j},A^{2}_{j},...A^{n}_{j})$. Because we consider symmetric
equilibria, ex-ante (before we condition on individual types), the individual
actions $A^{i}_{j}$ are random variables that are identically distributed
across $i\in{1,...,n}$. By the Central Limit Theorem, the distribution of each
vote total $V_{j}$ converges to a Normal distribution.

For each population size $n\in\mathbb{N}$, we use the \emph{Berry-Esseen theorem} (in the Appendix), 
to bound the size of the error in the Central Limit Theorem. As an immediate consequence of the Berry-Esseen theorem, 
for each $n$, we obtain bounds on the probability that the sum of
independent, identically distributed random variables (such as our
unconditional vote totals) take a value within a given interval $J\in
\mathbb{R}$ (Lemma \ref{lemma:anti-conc}, also in the Appendix). 
These bounds tighten with the variance of the independent, identically distributed
random variables, and in Lemma \ref{lemma:var-bound} (in the Appendix as well) we
establish a lower bound on this variance, and with it an upper bound on the
probability that the sum of vote totals lies within a given interval. We are
then ready for the key result of this subsection.

\begin{proposition}
\label{proposition:small-marg-piv} There is a sequence of real numbers
$\varepsilon_{n}\rightarrow0$ such that for any sample size $n\geq1$, any
Bayes-Nash equilibrium, and every pair $j,k$ of distinct alternatives,
\begin{equation}
\label{eq:small-marg-piv}E(Q_{j}Q_{k})\leq\varepsilon_{n}.
\end{equation}

\end{proposition}

We have established that unconditional marginal pivotalities vanish if the population
is sufficiently large. Since conditional pivotalities are bounded in
proportion to the unconditional ones (Lemma \ref{lemma:marg-piv}), it follows
that conditional pivotalities vanish as well. With vanishing marginal
pivotalities, the marginal benefit of taking a non-zero action vanishes as
well, and thus by the first-order condition (equation \ref{eq:foc-A}), it
follows that all individual actions converge to zero. It remains to be shown
that these individual actions converge to zero in proportion to each agent
$i$'s value for each action $j$, given by the realization $u^{i}_{j}$ of agent
$i$'s type.

\medskip\noindent\textbf{Step 3. Ratios of marginal pivotalities converge to
one}. \newline We noted above that agents with different type realizations
have different marginal pivotalities. In this subsection we establish that
these marginal pivotalities converge. First we derive two consequences from
Proposition \ref{proposition:small-marg-piv}: a unique alternative
asymptotically amasses almost all probability of being chosen (Corollary
\ref{corollary:likely-winner}), and the equilibrium is in pure strategies, as
the best responses are unique (Proposition \ref{proposition:pure-strategies}).

\begin{corollary}
\label{corollary:likely-winner} There is a sequence of real numbers
$\varepsilon_{n}\downarrow0$ such that for any sample size $n$ and every BN
equilibrium,
\[
P \left\{  \max_{j}Q_{j}\leq1-\varepsilon_{n} \right\}  \leq\varepsilon_{n}.
\]

\end{corollary}

We refer to the alternative that is chosen with probability converging to one
as the ``likely winner." Ultimately, we want to show that this likely winner is the
utilitarian optimum with probability converging to one. It suffices to show
that the likely winner is alternative 1. So downstream in this proof we will
establish that, indeed, the likely winner is alternative 1. For the time
being, we establish that the equilibrium is in pure strategies.

\begin{proposition}
\label{proposition:pure-strategies} For any $c>0$, there exists $n_{c}\in\mathbb{N}$
such that for all $n \geq n_{c}$, in any symmetric Bayes-Nash
equilibrium, for any type $u \in[0,1]^{m}$ there is only one best response
$\alpha(u)\in\mathcal{A}$.
\end{proposition}

We now reach a key step in our proof: establishing that the ratio of marginal
pivotalities converges to one across types (and across agents). If marginal
pivotalities were equal across types (or, if agents behaviorally misperceived
them to be equal), we could apply a heuristic logic reminiscent of Lalley and
Weyl (2019) or Goeree and Zhang (2017): with constant and equal marginal
pivotalities, the marginal benefit of increasing $a^{i}_{j}$ is constant in
$a^{i}_{j}$ but linear in $u^{i}_{j}$, while the marginal cost is linear in
$a^{i}_{j}$, so equating marginal benefit and marginal cost results in optimal
actions that are linear in $u^{i}_{j}$.

However, marginal pivotalities are not constant in actions: if, all else
equal, for a pair of agents $h$ and $i$, $a^{h}_{j}>a^{i}_{j}$, then agent
$i$'s expectation of $V_{j}$ is strictly greater than agent $j$'s, so their
expectations of the probability $Q_{j}$ that $j$ is chosen, and of the
likelihood that additional marginal support for $j$ is pivotal are different
as well. It follows that agents with different types face different incentives
\emph{insofar as they take different actions}. An important intuition is that
as marginal pivotalities and actions asymptotically vanish (Proposition
\ref{proposition:small-marg-piv}), the discrepancy in marginal incentives
vanishes as well, so marginal pivotalities across types converge, as marginal
actions across types converge.

We formalize this key convergence result in Proposition
\ref{proposition:coalescence-opinion}.

\begin{proposition}
\label{proposition:coalescence-opinion} There is a sequence of real numbers
$\delta_{n}\downarrow0$ such that for any sample size $n$, any BN equilibrium,
and each pair $j,k$ of distinct alternatives,
\begin{equation}
\label{eq:coalescence-opinion}(1+\delta_{n})^{-1}\leq\frac{E_{i}(Q_{j}Q_{k}%
)}{E(Q_{j}Q_{k})}\leq(1+\delta_{n})
\end{equation}
almost surely, for all agents $i=1,2,\cdots,n$.
\end{proposition}

\medskip\noindent\textbf{Step 4. Approximate equilibrium actions}. \newline
Proposition~\ref{proposition:coalescence-opinion} implies that, at least for
large sample sizes $n$, the \emph{marginal pivotalities} of the different
agents must be very nearly equal. Hence, by the first-order conditions
\eqref{eq:foc}, the votes $A^{i}_{j}$ are nearly linear functions of the
values $U^{i}$: in particular,
\[
2cA^{i}_{j} \approx\sum_{k\,:\, k\not =j} (U^{i}_{j}-U^{i}_{k})E(Q_{j}Q_{k}).
\]
The following lemma quantifies the size of the error in this approximation.

\begin{lemma}
\label{lemma:vote-bounds} Let $\delta_{n}$ be the sequence of real numbers
provided by Proposition~\ref{proposition:coalescence-opinion}. Then for each
agent $i$ and each alternative $j \in[m]$,
\begin{align}
\label{eq:upperv-b}2cA^{i}_{j}  &  \leq(1+\delta_{n})\sum_{k\,:\, k\not =j}
(U^{i}_{j}-U^{i}_{k})E(Q_{j}Q_{k}) +2\delta_{n}\sum_{k\,:\, k\not =j}%
E(Q_{j}Q_{k}),\\
\label{eq:lowerv-b}
2cA^{i}_{j}  &  \geq(1+\delta_{n})^{-1}\sum_{k\,:\, k\not =j} (U^{i}_{j}%
-U^{i}_{k})E(Q_{j}Q_{k}) -2 \delta_{n}\sum_{k\,:\, k\not =j}E(Q_{j}Q_{k}).
\end{align}

\end{lemma}

From the approximate individual actions given by Lemma \ref{lemma:vote-bounds}%
, we obtain as Corollary \ref{corollary:voteBounds} (in the Appendix) an
approximate aggregate support $V_{j}$ for each alternative $j$. The upper and
lower bounds of aggregate support in Corollary \ref{corollary:voteBounds} can
be expressed in an easier form. To do so, define
\[
\Delta:= \min_{1 \leq j \leq m-1} (\mu_{j}-\mu_{j+1})>0,
\]
and recall that by Assumption~\ref{assumption:F1} the means $\mu_{j}$ are
strictly ordered. Thus, $\Delta>0$. Then, we obtain the following bounds.

\begin{corollary}
\label{corollary:BetterVBounds} Let $\delta_{n}\downarrow0$ be the constants
provided by Proposition~\ref{proposition:coalescence-opinion}. Then for any
$\varepsilon\leq\Delta/4$, any sample size $n$, and any BN equilibrium, the
vote totals $V_{j}$ satisfy the inequalities
\begin{align}
\label{eq:v-upper-improved}2cV_{j}/n  &  \leq(1-3\Delta^{-1}\delta_{n}%
)\sum_{k=1}^{j-1} (\mu_{j}-\mu_{k}+ 2\varepsilon)E(Q_{j}Q_{k})\\
&  + (1+5\Delta^{-1}\delta_{n})\sum_{k=j+1}^{m} (\mu_{j}-\mu_{k}+
2\varepsilon)E(Q_{j}Q_{k}) \quad\text{and}\nonumber\\
\label{eq:v-lower-improved}
2cV_{j}/n  &  \geq(1+5\Delta^{-1}\delta_{n})\sum_{k=1}^{j-1}(\mu_{j}-\mu_{k}-
2\varepsilon)E(Q_{j}Q_{k})\\
&  + (1-3\Delta^{-1}\delta_{n})\sum_{k=j+1}^{m}(\mu_{j}-\mu_{k}-
2\varepsilon)E(Q_{j}Q_{k})\nonumber
\end{align}
except on an event of probability at most $2m \exp\left\{  -2 n \varepsilon
^{2} \right\}  $.
\end{corollary}

\medskip\noindent\textbf{Step 5. Asymptotic properties of all equilibria}.
\newline Notice that for alternative $j=1$, the lower bound in expression
\ref{eq:v-lower-improved} is strictly positive; whereas, for $j=m$, the upper
bound in expression \ref{eq:v-upper-improved} is negative.

\begin{corollary}
\label{corollary:extremes} For any $\varepsilon< \Delta/4$, for all
sufficiently large sample sizes $n$, in any BN equilibrium,
\begin{equation}
\label{eq:1}P \left\{  V_{m}<0<V_{1} \right\}  \geq1- 2m e^{-2 \varepsilon^{2}
n}.
\end{equation}

\end{corollary}

\bigskip\noindent\textbf{Note:} Setting $\varepsilon= \Delta/8$ we obtain the
bound
\begin{equation}
\label{eq:expBound2}P \left\{  V_{m}<0<V_{1} \right\}  \geq1- 2m
e^{-\Delta^{2} n/32}.
\end{equation}

It follows that alternative $1$ asymptotically receives more support than
alternative $m$. We can also show at this stage that while pivotality vanishes
to zero, it vanishes slowly, in the following precise sense.

\begin{corollary}
\label{corollary:bigQQ} There exist constants $\theta_{n}\rightarrow\infty$
such that for every sample size $n$ and any BN equilibrium,
\begin{equation}
\label{eq:bigQQ}n\max_{j}\max_{k\,:\,k \not = j} E(Q_{j}Q_{k})\geq\theta_{n}.
\end{equation}

\end{corollary}

We use the bounds on support in Corollary \ref{corollary:BetterVBounds}, the
finding that alternative $1$ has strictly greater support than $m$ (Corollary
\ref{corollary:extremes}), and the slow convergence result in Corollary
\ref{corollary:bigQQ}) repeatedly in the final steps of our proof. In these
steps, we establish that alternative $1$ is chosen with probability
arbitrarily close to one in sufficiently large collective entities, first for the easier
case with $m=2$ alternatives, then for the still simpler case with $m=3$
alternatives, and finally for the more elaborate case with $m>3$ alternatives.

Throughout the rest of the section, let

\begin{itemize}
\item $\delta_{n}\downarrow0$ be the constants in
Corollary~\ref{corollary:BetterVBounds},

\item $\varepsilon_{n}\downarrow0$ be the constants in
Corollary~\ref{corollary:likely-winner},

\item $\theta_{n}\uparrow\infty$ be the constants in
Corollary~\ref{corollary:bigQQ}, and

\item $\Delta=\min_{j\leq m-1}(\mu_{j}-\mu_{j+1})$.
\end{itemize}

In using Corollaries \ref{corollary:BetterVBounds} and
\ref{corollary:extremes}, we shall always take $\varepsilon= \Delta/8$; for
this choice the exceptional events in the two corollaries have probabilities
bounded by $2m e^{-\Delta^{2}n/32} $, and for any pair $j<k$ of alternatives,
\begin{align*}
(\mu_{j}-\mu_{k}+ 2\varepsilon)  &  \leq1+ \Delta/4 \quad\text{and}\\
(\mu_{j}- \mu_{k}- 2 \varepsilon)  &  \geq3\Delta/4.
\end{align*}

\medskip\noindent\textbf{Step 6a. Proof of Theorem \ref{theorem:main} for
$m=2$ alternatives}. \newline Since the selection probabilities $Q_{k}$ are
proportional to the exponentiated vote totals $e^{V_{k}}$,
Corollary~\ref{corollary:extremes} implies that $Q_{1}>Q_{2}$ except on an
event of probability no larger than $4 \exp\left\{  -\Delta^{2}n/32 \right\}
$. By Corollary~\ref{corollary:likely-winner}, $\max(Q_{1},Q_{2})$ exceeds
$1-\varepsilon_{n}$ with probability at least $1-\varepsilon_{n}$; hence,
\[
P \left\{  Q_{1}\leq1-\varepsilon_{n} \right\}  \leq4 e^{-\Delta^{2}%
n/32}+\varepsilon_{n} \longrightarrow0.
\]
Therefore, relation \eqref{eq:main} holds with $\beta_{n}=\varepsilon
_{n}+4e^{-\Delta^{2}n/32}$. This proves Theorem~\ref{theorem:main} in the case
of $m=2$ alternatives. \qed

\medskip\noindent\textbf{Step 6b. Proof of Theorem \ref{theorem:main} for
$m=3$ alternatives}. \newline Corollary~\ref{corollary:likely-winner} implies
that $\max_{j}Q_{j}\geq1-\varepsilon_{n}$ with probability at least
$1-\varepsilon_{n}$, and Corollary~\ref{corollary:extremes} implies that
$Q_{1}>Q_{3}$, and therefore $Q_{3}\not = \max_{j} Q_{j}$, with probability at
least $1-6\exp\left\{  -\Delta^{2}n/32 \right\}  $. Consequently, if relation
\eqref{eq:main} were not true then there would be some $\varrho>0$ such that
for infinitely many $n$, in some BN equilibrium,
\begin{equation}
\label{eq:badMax2}P \left\{  Q_{2}\geq1-\varepsilon_{n} \right\}  \geq\varrho.
\end{equation}
We want to show that this cannot be true.

The first step is to show that $E(Q_{2}Q_{3})$ must, for large $n$, be much
smaller than $E(Q_{2}Q_{1})$. By Corollary \ref{corollary:bigQQ}, for any
sample size $n$ and every BN equilibrium,
\[
n \max(E(Q_{1}Q_{2}), E(Q_{1}Q_{3}), E(Q_{2}Q_{3}))\geq\theta_{n}.
\]
By Corollary \ref{corollary:extremes} we have $V_{1}>0>V_{3}$, and therefore
$Q_{1}>Q_{3}$, except on an event $B$ of probability $P(B)\leq6 \exp\left\{
-\Delta^{2}n/32 \right\}  $; consequently,
\begin{align*}
E(Q_{2}Q_{3})  &  = E(Q_{2}Q_{3}\mathbf{1}_{B^{c}})+E(Q_{2}Q_{3}\mathbf{1}%
_{B})\\
&  \leq E(Q_{1}Q_{3}\mathbf{1}_{B})+ P(B)\\
&  \leq E(Q_{1}Q_{3})+ 6 \exp\left\{  -\Delta^{2}n/32 \right\}  .
\end{align*}
This implies that, for large enough $n$, either $n E(Q_{1}Q_{2})$ or
$nE(Q_{1}Q_{3})$ must exceed $\theta_{n}/2$, and so $nE(Q_{1}(1-Q_{1}%
))\geq\theta_{n}/2$. It follows, by the lower bound
\eqref{eq:v-lower-improved}, that for all sufficiently large $n$,
\begin{align*}
2cV_{1}  &  \geq n (1-3\Delta^{-1}\delta_{n})\sum_{j=2,3}(\mu_{1}-\mu
_{j}-2\varepsilon) E(Q_{1}Q_{j})\\
&  \geq n (1-3\Delta^{-1}\delta_{n})({\Delta}/{2})E(Q_{1}(1-Q_{1}))\\
&  \geq\Delta\theta_{n}/8
\end{align*}
except on an event $B^{\prime}$ of probability $\leq6 \exp\left\{  -\Delta
^{2}n/32 \right\}  $. (Here we have used the fact that $\delta_{n}%
\rightarrow0$ as $n \rightarrow\infty$.) Now $V_{3}<0$ except on the event
$B=\left\{  V_{1}\leq0 \right\}  \cup\left\{  V_{3}\geq0 \right\}  $,which has
probability at most $6 \exp\left\{  -\Delta^{2}n/32 \right\}  $; consequently,
except on the event $B^{\prime\prime}:=B\cup B^{\prime}$, whose probability is
at most $12 \exp\left\{  -\Delta^{2}n/32 \right\}  $,
\[
\frac{Q_{1}}{Q_{3}} =\frac{e^{V_{1}}}{e^{V_{3}}}> \exp\left\{  \Delta
\theta_{n}/(16 c) \right\}  .
\]
Thus,
\begin{align*}
E(Q_{2}Q_{3}\mathbf{1}_{B^{\prime\prime}})  &  \leq P(B^{\prime\prime})\leq12
\exp\left\{  -\Delta^{2}n/32 \right\}  \quad\text{and}\\
E(Q_{2}Q_{3}\mathbf{1}_{(B^{\prime\prime c}})  &  \leq E(Q_{2}Q_{1}%
\mathbf{1}_{(B^{\prime\prime c}}) \times\exp\left\{  -\Delta\theta_{n}/(16 c)
\right\}  ,
\end{align*}
which together show that
\begin{equation}
\label{eq:q2q3}E(Q_{2}Q_{3}) \leq E(Q_{2}Q_{1}) \times\exp\left\{
-\Delta\theta_{n}/(16c) \right\}  +12 \exp\left\{  -\Delta^{2}n/32 \right\}  .
\end{equation}

We next use the inequality \eqref{eq:q2q3} in conjunction with the upper bound
\eqref{eq:v-upper-improved} to show that the vote totals $V_{2}$ must be
bounded above by constants $\beta_{n}\rightarrow0$, except on events of
probability $\leq6 \exp\left\{  -\Delta^{2}n/32 \right\}  $, at least for
sufficiently large $n$. Since $P \left\{  V_{1}\leq0 \right\}  \leq6
\exp\left\{  -\Delta^{2}n/32 \right\}  $, by
Corollary~\ref{corollary:extremes}, it will then follow that the ratio
$Q_{2}/Q_{1}$ is bounded above by $e^{\beta_{n}}$ except for an event of
probability $\leq12 \exp\left\{  -\Delta^{2}n/32 \right\}  $, a contradiction
to the hypothesis \eqref{eq:badMax2}.

The details are as follows: by \eqref{eq:v-upper-improved} and
\eqref{eq:q2q3}, except on an event of probability no larger than $6
\exp\left\{  -\Delta^{2}n/32 \right\}  $,
\begin{align*}
2cV_{2}\leq &  - n (1-3 \Delta^{-1}\delta_{n})(\mu_{1}-\mu_{2}- \Delta
/4)E(Q_{1}Q_{2})\\
&  +n(1+5 \Delta^{-1}\delta_{n})(\mu_{2}-\mu_{3}+ \Delta/4)E(Q_{1}%
Q_{2})e^{-\Delta\theta_{n}/(16c)}\\
&  + n(1+5 \Delta^{-1}\delta_{n})(\mu_{2}-\mu_{3}+ \Delta/4) (12 \exp\left\{
-\Delta^{2}n/32 \right\}  ).
\end{align*}
Since $\theta_{n}\rightarrow\infty$ and $\delta_{n}\downarrow0$, the sum of
the first two terms is negative for all large $n $. Thus, for large $n $ we
have
\[
V_{2}\leq n \times2 \times(1+ \Delta/4) \times12\exp\left\{  -\Delta^{2}n/32
\right\}  /(2c) \longrightarrow0
\]
except on an event of probability $\leq6 e^{-\Delta^{2}n/32}$. By
Corollary~\ref{corollary:extremes}, $V_{1}>0$ except on an event of
probability $\leq6 \exp\left\{  -\Delta^{2}n/32 \right\}  $, and so for all
large $n$ we have $V_{2}-V_{1}\leq V_{2}\leq1$, and hence $Q_{2}/Q_{1}\leq
e^{1}$, except on an event of probability $\leq12 e^{-\Delta^{2}n/32}$. Thus,
\eqref{eq:badMax2} is impossible for large $n$. This proves
Theorem~\ref{theorem:main} in the case $m=3$.

\qed

\medskip\noindent\textbf{Step 6c. Proof of Theorem \ref{theorem:main} for
$m\geq3$ alternatives}. \newline In the case $m=3$ we were able to show
directly that $V_{1}$ is bounded below by a large constant $(\Delta\theta
_{n}/16c)$ except on an event of exponentially small probability by comparing
the expectations $E(Q_{1}Q_{2})$ and $E(Q_{2}Q_{3})$. This argument fails when
$m \geq4$. Nevertheless, it can be shown that the expectations $nE(Q_{1}%
Q_{j})$ cannot all remain bounded, as follows.

\begin{proposition}
\label{proposition:bigQ1} There is a sequence $\xi_{n}\uparrow\infty$ such
that for every sample size $n$, in every BN equilibrium,
\begin{equation}
\label{eq:bigQ1}\max_{j \geq2}nE(Q_{1}Q_{j})\geq\xi_{n}%
\end{equation}

\end{proposition}

Proposition \ref{proposition:bigQ1} strengthens Corollary
\ref{corollary:bigQQ}. Where Corollary \ref{corollary:bigQQ} noted that the
marginal pivotality of supporting some alternative vanishes slowly,
Proposition \ref{proposition:bigQ1} means that in particular the marginal
pivotality of supporting alternative $1$ vanishes slowly. Its proof is
cumbersome and we relegate it to the Appendix. Having established that
marginal pivotality for alternative $1$ vanishes slowly, the penultimate step
in our proof of Theorem \ref{theorem:main} for three or more alternatives is
to show that support for other alternatives is asymptotically bounded.

\begin{proposition}
\label{proposition:Vsmall} For all sufficiently large $n$, in any BN
equilibrium,
\begin{equation}
\label{eq:conclusion-c2}P \left\{  \max_{j \geq2}V_{j}>1 \right\}  \leq2m^{2}
\exp\left\{  - \Delta^{2}n/32 \right\}  .
\end{equation}

\end{proposition}

Since in large population, almost all probability of being chosen is
concentrated in a single alternative (Corollary \ref{corollary:likely-winner},
it is then intuitive that this alternative almost certain to be chose is the
one not constrained by these low upper bounds, i.e. alternative one. This is
the last step of our proof.
\begin{proof} [Completion of Proof of Theorem \ref{theorem:main} for $m\geq3$ alternatives]
By Corollary~\ref{corollary:extremes}, $V_{1}>0$ except on an event
of probability $\leq 2m \exp \xset {-\Delta^{2}n/32} $, and by
Proposition~\ref{proposition:Vsmall}, $\max_{ j\geq 2}V_{j}\leq 1$
except on an event of probability $\leq 2m^{2} \exp \xset
{-\Delta^{2}n/32} $, provided $n$ is sufficiently large. Therefore,
for all large $n$,
\begin{displaymath}
P \xset {\max_{j \geq 2}\frac{Q_{j}}{Q_{1}}\geq e^{1}} \leq
2m(m+1)\exp \xset {-\Delta^{2}n/32} .
\end{displaymath}
But by
Corollary~\ref{corollary:likely-winner},
\begin{displaymath}
P \xset {\max _{1 \leq j \leq m}Q_{j}\leq 1- \varepsilon _{n}}\leq \varepsilon_{n};
\end{displaymath}
consequently, for all $n$  large enough that $(1-
\varepsilon_{n})/\varepsilon_{n}> e^{1}$,
\begin{align*}
P \xset {Q_{1}\leq 1-\varepsilon_{n}}&\leq P \xset
{\max_{1 \leq j \leq m} Q_{j}\leq 1-
\varepsilon_{n}}+P \xset
{\max_{j\geq
2}Q_{j}\geq 1-\varepsilon_{n}}\\
&\leq \varepsilon_{n}+P \xset
{\max_{j \geq 2}(Q_{j}/Q_{1})>e^{1}}\\
& \leq \varepsilon_{n}+  2m(m+1)\exp \xset {-\Delta^{2}n/32} .
\end{align*}
\end{proof}

\section{Discussion\label{sec:disc}}

We have established that in a collective multiple choice problem, a Quadratic
Transfers Mechanism (QTM) chooses the utilitarian welfare-maximizing choice
with probability converging to one in the size of the population. Under QTM, 
agents face a quadratic menu of costs of expressing support for
or opposition to each alternative. Given these costs, an agent's support for
a given alternative increases---holding all else fixed---in the alternative's value to the agent. 
The magnitude of individual expressions of support or opposition decreases
in the size of the population, but it decreases slowly, 
so the magnitude of aggregate support or opposition for each alternative diverges to infinity. 
Hence, any individual deviation to any affordable action plays, asymptotically, no role at all;
that is, QTM still chooses the utilitarian optimum even if an agent 
(or a finite subset of agents) arbitrarily departs from equilibrium play. 

We illustrate through a numerical example that convergence to the desired
outcome can happen for a relatively small group.  

\begin{example} \emph{Consider a group with $n=300$ agents facing a collective choice over three alternatives ${1,2,3}.$ Each agent's type (vector of values for the three alternatives) is either $(3,2,0)$ with probability $p > \frac{1}{2}$, or $(0,2,3)$ with probability $1-p$.
Alternative $2$ is the unique utilitarian choice if between $101$ and $199$ agents are of each type, which occurs with probability 1 (up to eight decimal digits) when the group is expected to be almost evenly divided (i.e. when $p \rightarrow \frac{1}{2}$). 
In such cases, under the QTM mechanism alternative $2$ is chosen in equilibrium with probability greater than 96\%. Whereas, under voting with plurality rule and full turnout, alternative $2$ is chosen with probability $0$ in equilibria in which every agent votes sincerely for her preferred alternative.}
\end{example}

We now set to explore the robustness of our result to alternative informational environments. 

First, in early versions of this paper, we considered a complete information
environment, in which the matrix $u$ of agents' values over alternatives is 
common knowledge among agents (while still unknown to the mechanism), 
and the solution concept is Nash equilibrium. Our result is robust in such environment: with probability converging to one in the size of the population, the utilitarian optimum is chosen by QTM.\footnote{Rigorous formal arguments backing up this claim are available from the authors. Palfrey and Rosenthal (1985) argue that the complete information environment is more suitable for the modelling of collective choice in committees, while incomplete information fits better to the analysis of decision-making in large groups.}

Second, we turn attention to the assumption that the type generating process is known to the voters. This assumption is common in the literature on the welfare consequences of alternative decision-making procedures in private values contexts, including standard voting rules (e.g. Palfrey 1989, Ledyard and Palfrey 2002, Myerson 2002, Bouton 2013), systems that allow vote transfers between issues or voters (e.g. Casella 2005, Casella et al. 2012), and elections with costly participation (e.g. B\"{o}rgers 2004, Herrera et al. 2014, Krishna and Morgan 2015, Gr\"{u}ner and Tr\"{o}ger 2019). Therefore, our paper is no exception in this regard. However, this assumption is not innocuous (see Wilson 1987), and we seek to understand the limitations that it imposes on our results.

Assume that agents not only do not know the matrix $u$ of values over alternatives, but further, they also ignore the distribution $F$ from which such values are drawn; instead, they only have prior beliefs about such distribution, and these beliefs may vary across agents. Our result does not survive in the most general environment in which beliefs about the distribution of preferences correlate with actual preferences: if agents who prefer a collectively suboptimal alternative (incorrectly) believe that they are more pivotal than they truly are, they will take actions of greater magnitude in support of their preferred alternative, potentially overturning---if there are enough of them, and their beliefs are biased enough---the efficient choice.\footnote{Similar challenges arise if types are correlated: as it is often the case in collective choice problems, optimality results that hold for independently-drawn types are not robust to type correlation (see among others, Bognar, Börgers and Moritz Meyer-ter-Vehn 2015; or Drexl and Kleiner 2018).}
 
On the other hand, if preferences are uncorrelated with beliefs about the distribution of preferences, then our optimality result is robust to some generalizations. Assume, for instance, that the true distribution of values over alternatives is drawn from $F$, and indeed a subset of agents believe it to be common knowledge that the distribution is $F$, but an arbitrary subset of group believe instead that it is common knowledge that that the distribution is in fact $G$. Regardless of the size of the set of agents with wrong beliefs, if there are only two alternatives ($m=2$), or, regardless of the number of alternatives, if the alternative that is in expectation optimal is the same under distributions $F$ and $G$, then the efficient alternative is chosen with asymptotic probability one. If $m=2$, regardless of their beliefs about the distribution of preferences, agents' actions converge to linear in their net value of alternative one minus the value of alternative two, so integrating over the true distribution, the net support for the optimal alternative is positive, while the net support for the suboptimal alternative is negative, for both the subpopulation with correct beliefs, and the subpopulation with wrong beliefs.
Further, for any $m$, if alternative $1$ is in expectation the optimal one both under $F$ and under $G$, then the subpopulation with wrong beliefs expects $1$ to be chosen, and subject to being pivotal, they expect that their pivotality would be about choosing $1$ or choosing something else; on net, they prefer $1.$

The collective entity may also choose optimally in some environments with multiple alternatives in which agents believe (incorrectly) that a suboptimal alternative is in fact optimal. Suppose that the vector of expected values according to the true distribution $F$ is $(\alpha,\beta,\mu_{3},...,\mu_{m})$, but a fraction of the group believe that it is common knowledge that the distribution is in fact $G$ with vector of expected values $(\beta,\alpha,\mu_{3},...,\mu_{m})$, with $\alpha>\beta$ and $\beta-\mu_{j}>\alpha-\beta$ for any $j\in\{3,4,...,m\}.$ For any size of the subpopulation with wrong beliefs, and for any sequence of equilibria in which alternatives $1$ and $2$ receive the most and second-most support, the truly optimal alternative $1$ is chosen with probability asymptotically one. The intuition here is that agents with wrong beliefs expect $2$ to win, but they expect that if their action is pivotal, it is most likely to be pivotal between $1$ and $2$, and conditional on this event, aggregating according to the true distribution they on net prefer $1$ so they'll vote on net for $1.$\footnote{In fact, return to Example 1, and assume that while the true distribution assigns almost equal probability to the two types (i.e. $p \rightarrow \frac{1}{2}$), all agents mistakenly believe that $p$ is substantially larger than $\frac{1}{2}$. If agents believe that $p>\frac{2}{3}$, then agents also believe that alternative $3$ is the efficient one. And yet, we find numerically that with such mistaken beliefs, the truly optimal alternative $2$ is still the most likely winner (e.g. when $n=300$ the win probability of the truly optimal alternative is above 57\%, even in the extreme case in which $p\rightarrow \frac{1}{2}$ but agents incorrectly believe that $p$ is close to $1$).} 

Our optimality results rely on further stylized assumptions that are standard in the
theoretical literature on mechanism design and implementation, including:
agents' rationality, agents' individual optimization without collusion, and quasilinear preferences (the list is not exclusive). We comment specifically on quasilinearity.

We have assumed that each agent's type or value $u^{i}$ is a vector that
characterizes agent $i$'s ordinal preferences over (wealth, alternative) pairs
and over lotteries of such pairs, so that for any two (wealth, alternative)
pairs $(w,j)$ and $(w^{\prime},k)$, agent $i$ prefers $(w,j)$ to $(w^{\prime
},k)$ if and only if
\begin{equation}
\label{ineq:risk-neut}w+u^{i}_{j} \geq w^{\prime i}_{k}%
\end{equation}
and for any pair of lotteries ${p,q}$ respectively generating the random
(wealth, chosen alternative) vectors $(W_{p},C_{p})$ and $(W_{q},C_{q})$,
agent $i$ prefers lottery $p$ over lottery $q$ if the expected value of
$W_{p}+U^{i}_{C_{p}}$ is greater than the expected value of $W_{q}%
+U^{i}_{C_{q}}$. The linearity over wealth implies we are assuming that agents
are risk-neutral. Under this restriction to risk neutrality, the value
$u^{i}_{j}$ can be interpreted as agent $i$'s "willingness to pay" for
alternative $j$ to be chosen, where such willingness to pay is invariant on
agent $i$'s wealth. Since willingnesses to pay can be compared across agents,
in this purely ordinal environment we can interpret utilitarianism as
the normative principle of choosing the alternative with greatest aggregate
willingness to pay.

However, in most applications of interest, agents are risk-averse, and their
willingness to pay for an alternative to be chosen increases with wealth.
Consider an environment with risk-averse, expected utility maximizing agents
with separable preferences over wealth and over the alternative chosen, whose
preferences over lotteries over (wealth, chosen alternative) pairs are such
that there exists a concave real-valued function $g$ defined on$~\mathbb{R_{+}%
}$ such that $i$ prefers lottery $p$ that generates the random (wealth, chosen
alternative) vector $(W_{p},C_{p})$ over lottery $q$ that generates the random
vector $(W_{q},C_{q})$ if and only if the expected value of $g(W_{p}%
)+U^{i}_{C_{p}}$ is greater than the expected value of $g(W_{q})+U^{i}_{C_{q}%
}.$ \footnote{Kazumura, Mishra and Serizawa (2020) study dominant strategy
incentive compatibility in this environment, and they survey the nascent
literature on mechanism design without quasilinearity. Eguia and Xefteris
(2021) analyze vote-buying mechanisms for binary setups when the agent's
utility over wealth is concave. See, also, Casella and Mace (2021).}

Our main result (Theorem \ref{theorem:main}) applies but must be reinterpreted
in this environment with risk-averse agents, as follows. Enrich our benchmark
environment, endowing each agent $i$ with wealth $w^{i}$ as a primitive, in
addition to her vector of values $u^{i}$, and consider that $w$ units of
wealth give to the agent $g(w)$ units of utility, with $g$ being strictly
increasing, concave and differentiable. Because individual actions converge to
zero (an immediate consequence of Proposition \ref{proposition:small-marg-piv}%
), agent $i$'s final wealth converges to $w^{i}$ in all equilibria and for all
realizations of her vector of values. If agent $i$ estimates her utility over
wealth near $w^{i}$ using the local linear approximation of $g$, given by
$g(w)\approx g(w^{i})+g^{\prime}(w^{i})(w-w^{i})$, then agent $i$ prefers (wealth,
alternative) pair $(w,j)$ to pair $(w^{\prime},j^{\prime})$ if and only if
\[
g^{\prime}(w^{i})(w-w^{i})+u^{i}_{j}\geq g^{\prime }(w^{i})(w^{\prime}-w^{i})+u^{i}%
_{j^{\prime}},
\]
or, equivalently,
\begin{equation}
\label{eq:risk-av}\ w+u^{i}_{j}/g^{\prime}(w^{i})\geq w^{\prime}+u^{i}_{j'}/g^{\prime}(w^{i}),
\end{equation}

A group of agents with preferences represented by the inequality
(\ref{eq:risk-av}) is one that fits within our original framework, with value
types $\Tilde{u}^{i}:=u^{i}/g^{\prime}(w^{i})$. These values $\Tilde{u}^{i}$ still
represent willingness to pay, but now the primitive values $u^{i}$ are
distorted by wealth, so that agents with greater wealth $w^{i}$, and thus with
lower marginal utility of wealth $g^{\prime}(w^{i})$, have a greater willingness
to pay $\Tilde{u}^{i}_{j}$ for alternative $j$.

Theorem \ref{theorem:main} thus applies, and confirms that if the population is
sufficiently large, QTM chooses the alternative with the greatest aggregate
marginal willingness to pay with probability arbitrarily close to one. The
alternative chosen under QTM is the only Pareto optimal choice: for any other
choice, there is a set of wealth transfers such that executing these transfers
and choosing the alternative chosen by QTM makes every agent better off.

On the other hand, the normative appeal of the result under risk aversion is
weakened. First, because QTM aggregates (marginal) willingness to pay, and
risk-averse agents' willingness to pay depends on wealth, the QTM mechanism in
an environment with risk-aversion violates Walzer's (1983) normative principle
of separation of the spheres of economic and political power. According to
this principle, the distribution of power in one of these spheres should have no
influence over power in the other sphere. Specifically, because willingness to
pay increases in wealth, the preferences of wealthier agents are overweighed
in the aggregation of preferences, which introduces concerns about equity.
And, relatedly, the implementation of the utilitarian optimum (i.e. the
alternative that maximizes the sum of individual valuations) which takes place
under quasilinearity, is not guaranteed with risk aversion: the alternative
that maximizes aggregate marginal willingness to pay does not always coincide
with the one that maximizes the sum of values.

For these reasons, we interpret Theorem \ref{theorem:main} as a positive
result that contributes towards a better understanding of collective choice mechanisms with transfers, and not as the normative basis for a prescriptive recommendation for
the design of electoral systems or legislative procedures in practical applications.

\newpage 

\section*{Appendix \label{sec:appendix}}

For Online Publication. 

\begin{proof}[Proof of Proposition~\ref{proposition:existence}]
For any $n\in
\mathbb{N}
,$ consider the ex-ante version of our game, where each player selects a
distributional strategy, in the sense of Milgrom and Weber (1985). Let $
\mathcal{B}$ denote the\ Borel $\sigma -$algebra over $[0,1]^{m}\times
\mathcal{A}$. A distributional strategy\ for player $i$ is a
probability measure over $\mathcal{B}$ whose marginal distribution
over $[0,1]^{m}$ is $F.$ Denote this game $\Gamma _{MW(n)}.$  

Next, consider a normal form game $\Gamma ^{2}(n),$ in which the pure
strategy set of each agent is the set of all distributional strategies in
game $\Gamma _{MW(n)},$ and the payoff for each such strategy is equal to
the expected payoff of the corresponding distributional strategy. 
As noted in the proof of
Thm 1 of Milgrom and Weber (1985), page 626, when endowed with the weak
convergence topology, the sets of distributional strategies in game $\Gamma
_{MW(n)}$ are compact, so the set of pure strategies in game $\Gamma ^{2}(n)$
is compact. Game $\Gamma ^{2}(n)$ is symmetric, and hence it satisfies the
weaker condition of quasi-symmetry in the sense of Reny (1999). As indicated
in the proof of Thm 1 of Milgrom and Weber, page 626, the payoff function in
game $\Gamma _{MW(n)},$ and hence with it the payoff function in game $%
\Gamma ^{2}(n)$ is linear, and thus weakly quasiconcave, satisfying as well
the weaker condition of diagonal-quasiconcavity in Reny\ (1999). Because the
payoff function in game $\Gamma ^{2}(n)$ is also continuous, it satisfies
upper semi-continuity and Reny's conditions of diagonal payoff security. It
follows that game $\Gamma ^{2}(n)$ possesses a symmetric pure Nash
equilibrium (Reny 1999, Corollary 4.3). 
The distributional strategy profile selected in this symmetric pure strategy equilibrium of game $\Gamma ^{2}(n)$
is a symmetric equilibrium in distributional strategies of game $\Gamma _{MW(n)}$.
The behavioral strategy profile corresponding to this symmetric equilibrium in distributional strategies is a symmetric BN equilibrium of our game. 
\end{proof}

\begin{proof}[Proof of Lemma ~\ref{lemma:marg-piv}]
By equation \eqref{eq:rjka}, for any pair $j,k$ of alternatives and every $i
\in[n]$,
\begin{displaymath}
E_{i}(Q_{j}Q_{j})=r_{j,k}(A^{i}) \quad \textrm {where} \quad
r_{j,k}(a)=E(Q^{1}_{j}(a)Q^{1}_{k}(a)).
\end{displaymath}
Since the random vectors $A^{i}$ are independent and identically
distributed, the distribution of $E_{i}(Q_{j}Q_{k})$ is identical to
that of $E_{1}(Q_{j}Q_{k})$;  consequently, it suffices to prove
\eqref{eq:marg-piv} for $i=1$.
By hypothesis, the  vectors $A^{i}$ take value in the action
space $\mathcal{A}:=[-c^{-1/2},c^{-1/2}]^{m}$, and so the components
are bounded by $1/ \sqrt {c}$ in absolute value. Hence, by
definition \eqref{eq:Qija} of the random variables $Q_{j}(a)$, for any two
possible vote vectors $a,a^{*}\in \mathcal{A}$,
\begin{displaymath}
e^{-16/ \sqrt {c}}\leq
\frac{Q^{1}_{j}(a)Q^{1}_{k}(a)}{Q^{1}_{j}(a^{*})Q^{1}_{k}(a^{*})}
=\frac{r_{j,k}(a)}{r_{j,k}(a^{*})}\leq     e^{16/ \sqrt {c}}.
\end{displaymath}
Since $E(Q_{j}Q_{k})=Er_{j,k}(A^{1})$, the inequalities
\eqref{eq:q-marg-piv} follow.
\end{proof}

\begin{berryEsseenTh}
There is a constant $C_{BE} \leq3$ such that if $Y_{1},Y_{2}, \cdots$ are
independent, identically distributed random variables satisfying
\begin{equation}
\label{eq:berryEsseenTh-hyp}EY_{1}=0, \quad EY_{1}^{2}=\sigma^{2}>0,
\quad\text{and} \; E|Y_{1}|^{3}=\gamma<\infty,
\end{equation}
then for all $x\in\mathbb{R}$ and all integers $n \geq1$,
\begin{equation}
\label{eq:berryEsseenTh}\bigg\vert P \left\{  \frac{1}{\sigma\sqrt{n}}%
\sum_{i=1}^{n}Y_{i} \leq x \right\}  - \Phi(x) \bigg \vert \leq C_{BE}%
\frac{\gamma}{\sigma^{3}\sqrt{n}},
\end{equation}
where $\Phi$ denotes the standard normal cumulative distribution function.
\end{berryEsseenTh}

See Feller (1971), section XVI.5 for the proof. It is now known that
$C_{BE}\leq.5$: see Korolev \& Shevtsova (2010). Following is an immediate consequence.

\begin{lemma}
\label{lemma:anti-conc} For any $n\in\mathbb{N}$, if $Y_{1},Y_{2}, \cdots,Y_{n}$ are independent,
identically distributed random variables with variance $EY_{1}^{2}=\sigma
^{2}>0$, and absolute third moment $E|Y_{1}|^{3}<\infty$ then for every
interval $J \subset\mathbb{R}$ of positive finite length $|J|$,
\begin{equation}
\label{eq:anti-conc}P \left\{  \sum_{i=1}^{n}Y_{i}\in J \right\}  \leq
\frac{|J|}{\sigma\sqrt{2\pi n}}+ \frac{\gamma}{\sigma^{3}\sqrt{n}}.
\end{equation}
where $\gamma:=E|Y_{1} -EY_{1}|^{3}$.
\end{lemma}

\begin{proof}
First, if $Y_1$  has finite third moment then it has
finite first and second moments, by Jensen's inequality, so the mean $\mu:=EY_1$
is well-defined and finite. Now, if $J=[J_{low},J_{high}]$ and $J^n=[J_{low}-n EY_1,J_{high}+n EY_1]$, we observe that
\begin{center}
$P \xset{ \sum_{i=1}^n Y_i \in J } =  P\xset{ \sum_{i=1}^n (Y_i - EY_i ) \in J^n}$
\end{center}
\noindent
since the translated interval $J^n$ has the same length as $J$.
Moreover, given that the random variable $(Y_i - EY_i )$ has the same variance and third central moment
as the random variable $Y_i$, the result for a random variable with arbitrary mean follows from the result for a random variable with mean zero.
For this reason, for the rest of the proof we consider $EY_{1}=0$. Write $J=(x,y]$ and set $\sigma^{2}=\var
(Y_{1})$. By the Berry-Esseen theorem,  together with the
bound  $C_{BE}\leq .5$, we have
\begin{align*}
P \xset{ \sum_{i=1}^{n}Y_{i}\in J}&\leq \Phi (\frac{y}{\sigma\sqrt{n}})-\Phi
(\frac{x}{\sigma\sqrt{n}}) + 2C_{BE}\gamma /(\sigma^{3}\sqrt{n}) \\
&\leq \frac{|J|}{\sqrt{2 \pi n}\sigma} +\gamma /(\sigma^{3}\sqrt{n})
\end{align*}
the last because the standard normal density is bounded above by  $1/\sqrt{2
\pi}$.
\end{proof}

The usefulness of the bound \eqref{eq:anti-conc} is limited to circumstances
where upper bounds on the quantities $1/\sigma$ and $\gamma/\sigma$ are
available. To obtain such bounds, we will rely on the following elementary
sampling principle.

\begin{lemma}
\label{lemma:var-bound} Let $X$ be a random variable with finite second moment
such that for some real number $\varepsilon>0$
\begin{align}
\label{eq:variation-upper}P \left\{  X \leq-\varepsilon\right\}   &  \geq p >0
\quad\text{and}\\
P \left\{  X \geq+\varepsilon\right\}   &  \geq q >0.\nonumber
\end{align}
Then
\begin{equation}
\label{eq:variance-bound}\text{\textrm{var}} (X)\geq\varepsilon^{2} \min(p,q).
\end{equation}

\end{lemma}

\begin{proof}
Since $X$ has finite second moment, its mean $\mu$ and variance
$\var (X)$ are both well-defined and finite. Clearly, if $\mu \geq
0$ then $|\mu- (-\varepsilon )|\geq \varepsilon$, while if $\mu \leq
0$ then $|\mu- \varepsilon |\geq \varepsilon$. Hence,
\begin{displaymath}
P \xset {|X-\mu | \geq \varepsilon }\geq \min (p,q).
\end{displaymath}
The estimate \eqref{eq:variance-bound} now follows from the
definition $\var (X):= E(X-\mu)^{2}$ of variance.
\end{proof}

\begin{proof}[Proof of Proposition~\ref{proposition:small-marg-piv}]
By contradiction. Suppose the assertion is false; then for some
$\varepsilon >0$ and some pair $a,b$ of alternatives there will be
indefinitely large sample sizes $n$ such that for some BN
equilibrium, $E(Q_{a}Q_{b})>e^{16/\sqrt {c}}\varepsilon$. Consequently, by
Lemma~\ref{lemma:marg-piv}, in any such equilibrium every agent $i$
will have marginal pivotality $E_{i}(Q_{a}Q_{b})> \varepsilon$.
According to Assumption \ref{assumption:F2}, there are points $t_{a}e_{a}$ and
$t_{b}e_{b}$ such that $F$ assigns positive probability to every
neighborhood of $t_{a}e_{a}$ and $t_{b}e_{b}$. Fix $\varrho >0$
small enough that
\begin{equation}
\label{eq:rho-small}
(t_{*}-2 \varrho)\varepsilon - \varrho  \geq \varepsilon t_{*}/2
\quad \textrm {where} \; \; t_{*}:= \min (t_{a},t_{b}),
\end{equation}
and choose neighborhoods $G_{a},G_{b}$ of $t_{a}e_{a}$ and $t_{b}e_{b}$
such that for every point $x \in G_{k}$, where $k=a$ or $k=b$,
\begin{equation}
\label{eq:nhood-spec}
x_{k}\geq  t_{k} -\varrho \quad \textrm{and} \quad x_{j}<
\varrho \quad  \forall \; j \in [m]\setminus \xset {k}.
\end{equation}
(Here $[m]= \xset {1,2, \cdots ,m}$ is the set of alternatives.)
Denote by $\pi_{a}>0$ and $\pi_{b}>0$ the probabilities assigned to the
sets $G_{a}$ and $G_{b}$ by the sampling distribution $F$.
By the first-order
conditions for equilibria (equation \eqref{eq:foc}), on the event that agent $i$
has value $U^{i}\in G_{a}$ we have
\begin{align*}
2cA^{i}_{a} &= \sum_{k\,:\, k\not=a}
(U^{i}_{a}-U^{i}_{k})E_{i}(Q_{a}Q_{k}) \\
& \geq (t_{a}-2 \varrho) \sum_{k\,:\, k\not=a}E_{i}(Q_{a}Q_{k}) \\
& \geq (t_{a}-2 \varrho) E_{i}(Q_{a}Q_{b})\geq (t_{a}-2 \varrho) \varepsilon,
\end{align*}
and similarly,
\begin{align*}
2cA^{i}_{b} &=  \sum_{k\,:\, k\not=b}
(U^{i}_{b}-U^{i}_{k})E_{i}(Q_{b}Q_{k}) \\
&\leq (2 \varrho - t_{a})\varepsilon + \varrho \sum_{k\,:\, k\not=b} E_{i}(Q_{a}Q_{k})\leq -(t_{a}-2 \varrho )\varepsilon + \varrho,
\end{align*}
the last inequality because $Q_{b}\leq 1$ and $\sum_{k\in
[m]}Q_{k}=1$. The same argument shows that on the event that agent $i$ has value
$U^{i}\in G_{b}$ we have
\begin{align*}
2cA^{i}_{b} &\geq (t_{b}-2\varrho)\varepsilon  \quad \textrm
{and  }
2cA^{i}_{a} \leq -(t_{b}-2\varrho)\varepsilon  + \varrho.
\end{align*}
Thus, by \eqref{eq:rho-small},
\begin{align*}
2c(A^{i}_{a}-A^{i}_{b})& \geq \varepsilon t_{*} \quad \textrm {if} \;
U^{i}\in G_{a},  \\
2c(A^{i}_{a}-A^{i}_{b})&\leq -\varepsilon t_{*} \quad \textrm {if} \;
U^{i}\in G_{b}.
\end{align*}
Lemma~\ref{lemma:var-bound} now implies that
\begin{equation}
\label{eq:var-bound}
\var (A^{i}_{a}-A^{i}_{b}) \geq (\varepsilon t_{*}/2c)^{2} \min
(\pi_{a},\pi_{b}):= \sigma_{*}^{2}.
\end{equation}
Furthermore, since no agent will ever cast more than $1/ \sqrt {c}$
vote for or against any alternative, the absolute third central moment
satisfies
\begin{equation}
\label{eq:3dMomentBound}
E|A^{i}_{a}-A^{i}_{b}-E(A^{i}_{a}-A^{i}_{b})|^{3} \leq 8c^{-3/2}.
\end{equation}
According to our basic assumptions, the random variables
$(A^{i}_{a}-A^{i}_{b})_{i \in [n]}$ are independent and identically
distributed. Therefore, their sum $V_{a}-V_{b}$ is subject to the
conclusions of Lemma~\ref{lemma:anti-conc}; using this in conjunction
with the bounds \eqref{eq:var-bound} and \eqref{eq:3dMomentBound} for
the interval $J=[-n^{1/3},n^{1/3}]$ gives
\begin{displaymath}
P \xset {|V_{a}-V_{b}|\leq n^{1/3}}  \leq
\frac{2n^{1/3}}{\sigma_{*}\sqrt{2\pi n}}+
\frac{8c^{-3/2}}{\sigma_{*}^{3}\sqrt{n}}
\leq Cn^{-1/6}
\end{displaymath}
where $C=(2/\sigma_{*}\sqrt{2 \pi})+(8c^{-3/2}/ \sigma_{*})$. Since on the
event $\xset {|V_{a}-V_{b}|\geq n^{1/3}} $ one of the ratios
$Q_{a}/Q_{b}$ or $Q_{b}/Q_{a}$ must be at least
$\exp \xset {n^{1/3}}$, it  follows that
\begin{displaymath}
E(Q_{a}Q_{b})\leq  \exp \xset {-n^{1/3}} +C n^{-1/6} .
\end{displaymath}
This contradicts our hypothesis that  $E(Q_{a}Q_{b})>e^{16}\varepsilon$
holds for indefinitely large sample sizes $n$, and therefore completes
the proof of the proposition.
\end{proof}

\begin{proof} [Proof of Corollary ~\ref{corollary:likely-winner}]
Suppose not; then for some $\varepsilon >0$ there would be BN  equilibria
for indefinitely large $n$ such that
\begin{equation*}
P \xset {\max _{j}Q_{j}\leq 1-\varepsilon }\geq \varepsilon.
\end{equation*}
But if $\max _{j}Q_{j}\leq 1-\varepsilon$ then for some pair $j,k$  of
alternatives we would have $Q_{j}Q_{k}\geq  (1-\varepsilon)\varepsilon/m$,
and so
\begin{equation*}
E(Q_{j}Q_{k})\geq \varepsilon ^{2}(1- \varepsilon )/m
\end{equation*}
for infinitely many sample sizes $n$, contradicting Proposition  \ref%
{proposition:small-marg-piv}.
\end{proof}

\begin{proof} [Proof of Proposition ~\ref{proposition:pure-strategies}]
Recall (equation~\eqref{eq:foc-A}) that a necessary condition for a
best response $a$ for an agent $i$ with type $u^{i}$ is that
\begin{equation}
\label{eq:focB}
2ca_{j}= \sum _{k} (u^{i}_{j}-u^{i}_{k}) r_{j,k}(a)
\end{equation}
where $r_{j,k}(a)=EQ^{i}_{j}(a)Q^{i}_{k}(a)$. In any symmetric BN
equilibria, there must be at least one solution for any $u^{i}$ in the
support of the sampling distribution $F$. We will argue that if $n$ is
sufficiently large, then there is \emph{at most} one solution.
Suppose to the contrary that there were two distinct solutions
$a,a^{*}\in \mathcal{A}$, and let $j$ be an alternative for which
$|a_{j}-a^{*}_{j}|$ is maximal among all alternatives.
Then by the mean value theorem of calculus, there would exist a point
$a^{**}$ on the line segment from $a$ to $a^{*}$ such that
\begin{displaymath}
2c (a_{j}-a^{*}_{j}) = \sum _{k} (u^{i}_{j}-u^{i}_{k}) \nabla
r_{j,k}(a^{**}) \cdot (a-a^{*}),
\end{displaymath}
where $\nabla$ denotes the gradient operator and $\cdot$ denotes inner
product of vectors.  (Note: The action space $\mathcal{A}$ is convex,
so $a^{**}\in \mathcal{A}$.) By the choice of alternative $j$, it
follows that
\begin{gather}
\notag
2c \xnorm{a-a^{*}}/m \leq 2c |a_{j}-a^{*}_{j}|  \leq \sum_{k}
\xnorm{\nabla r_{j,k}(a^{**})} \xnorm{a-a^{*}} \quad \Longrightarrow
\\
\label{eq:grad-big}
\sum_{k}\xnorm{\nabla r_{j,k}(a^{**})}\geq 2c/m,
\end{gather}
where $\xnorm{ \cdot}$ denotes vector norm. Now the partial
derivatives of $r_{j,k}(a^{**})$ can be computed using the definition
\eqref{eq:rjka}: they are
\begin{align*}
\frac{\partial r_{j,k}}{\partial a^{**}_{\ell }}&=
-2E(Q^{1}_{j}(a^{**})Q^{1}_{k}(a^{**})Q^{1}_{\ell
}(a^{**})) \quad
\textrm {for}\;\;
\ell \not=j,k; \\
\frac{\partial r_{j,k}}{\partial
a^{**}_{k}}&=E(Q^{1}_{j}(a^{**})Q^{1}_{k}(a^{**}))-2E(Q^{1}_{j}(a^{**})Q^{1}_{k}(a^{**})^{2}).
\end{align*}
Since $0< Q^{1}_{j'}(\cdot )< 1$ for every $j'$, it follows that for
every triple $j,k,\ell $,
\begin{displaymath}
\left| \frac{\partial r_{j,k}}{\partial a^{**}_{\ell }} \right| \leq
3 \max_{j',k'}r_{j',k'}(a^{**}) \leq 3 e^{16/ \sqrt {c}}\max_{j',k'} E(Q_{j'}Q_{k'}),
\end{displaymath}
the last by Lemma~\ref{lemma:marg-piv}. But
Proposition~\ref{proposition:small-marg-piv} implies that
$\max_{j',k'} E(Q_{j'}Q_{k'})\leq \varepsilon_{n}\downarrow 0$, so
for large $n$ the inequality \eqref{eq:grad-big} is impossible.
\end{proof}

\begin{proof} [Proof of Proposition ~\ref{proposition:coalescence-opinion}]
By Proposition~\ref{proposition:small-marg-piv}, the average  marginal
pivotalities satisfy $E(Q_{j}Q_{k})\leq  \varepsilon_{n}$. Consequently, by
the first-order conditions  \eqref{eq:foc} and Assumption~\ref{assumption:F1}%
, the votes $A^{i}$  satisfy
\begin{equation*}
|A^{i}_{j}|\leq m \varepsilon_{n} /(2c):=\varepsilon^{\prime }_{n}.
\end{equation*}
By the same argument as in the proof of Lemma~\ref{lemma:marg-piv}, it
follows that for any $a,a^{*}\in\mathcal{A}$,
\begin{equation*}
e^{-16 \varepsilon ^{\prime }_{n}}\leq \frac{E((Q_{j}Q_{k})\,|\, A^{i}=a)}{%
E((Q_{j}Q_{k})\,|\, A^{i}=a^{*})} \leq e^{16 \varepsilon ^{\prime }_{n}}.
\end{equation*}
Hence, inequality \eqref{eq:coalescence-opinion} holds with  $%
1+\delta_{n}=\exp \xset {8 \varepsilon'_{n}}$.
\end{proof}

\begin{proof} [Proof of Lemma ~\ref{lemma:vote-bounds}]
Using the upper and lower bounds \eqref{eq:coalescence-opinion} on
the marginal pivotalities in conjunction with the first-order
conditions \eqref{eq:foc}, we obtain
\begin{align*}
2cA^{i}_{j}&\leq (1+\delta_{n})\sum _{k\,:\, k\not=j} (U^{i}_{j}-U^{i}_{k})_{+}
E(Q_{j}Q_{k})- (1+\delta_{n})^{-1}\sum _{k\,:\, k\not=j} (U^{i}_{j}-U^{i}_{k})_{-}
E(Q_{j}Q_{k})\\
&= (1+\delta_{n})\sum _{k\,:\, k\not=j} (U^{i}_{j}-U^{i}_{k})
E(Q_{j}Q_{k}) + ((1+\delta_{n})-(1+\delta_{n}^{-1}))\sum _{k\,:\, k\not=j} (U^{i}_{j}-U^{i}_{k})_{-}
E(Q_{j}Q_{k}) \\
&\leq (1+\delta_{n})\sum _{k\,:\, k\not=j} (U^{i}_{j}-U^{i}_{k})
E(Q_{j}Q_{k}) + 2\delta_{n}\sum _{k\,:\, k\not=j} (U^{i}_{j}-U^{i}_{k})_{-}
E(Q_{j}Q_{k}) \\
&\leq (1+\delta_{n})\sum _{k\,:\, k\not=j} (U^{i}_{j}-U^{i}_{k})
E(Q_{j}Q_{k}) + 2\delta_{n}\sum _{k\,:\, k\not=j} E(Q_{j}Q_{k}) .
\end{align*}
Here the subscripts $\pm$ denote positive and negative parts,
respectively, as defined at the end of
section~\ref{sec:model-assumpt-main}, and we have used the
elementary inequality
\begin{displaymath}
(1+ \delta)-(1+\delta)^{-1}\leq 2 \delta.
\end{displaymath}
This proves the upper bound  \eqref{eq:upperv-b}. The lower bound
\eqref{eq:lowerv-b} can be proved in similar fashion.
\end{proof}

We use Hoeffding's inequality (1963) to prove Corollary
\ref{corollary:voteBounds} below.

\begin{HoeffdingInequality}
Let $X_{1},X_{2},\cdots, X_{n}$ be independent random variables satisfying $0
\leq X_{i}\leq1$, and let $S_{n}=\sum_{i=1}^{n}X_{i}$ be their sum. Then for
any real $t>0$,
\begin{align}
\label{eq:hoeffding-ineq}P\left\{  |S_{n}- E S_{n}| \geq t \right\}   &  \leq2
e^{-2t^{2}/n} \quad\Longrightarrow\\
P\left\{  |S_{n}- E S_{n}| \geq nt \right\}   &  \leq2 e^{-2nt^{2}}.
\end{align}

\end{HoeffdingInequality}

\begin{corollary}
\label{corollary:voteBounds} Let $\delta_{n}\downarrow0$ be the constants
provided by Proposition~\ref{proposition:coalescence-opinion}. Then for any
$\varepsilon>0$, any sample size $n$, and any BN equilibrium, the vote totals
$V_{j}$ satisfy the inequalities
\begin{align}
\label{eq:v-upper}2cV_{j}/n  &  \leq(1+\delta_{n})\sum_{k\,:\, k\not =j}%
(\mu_{j}-\mu_{k}+ 2\varepsilon)E(Q_{j}Q_{k}) + 2\delta_{n} \sum_{k\,:\,
k\not =j} E(Q_{j}Q_{k}) \quad\text{and}\\
\label{eq:v-lower}
2cV_{j}/n  &  \geq(1+\delta_{n})^{-1}\sum_{k\,:\, k\not =j}(\mu_{j}-\mu
_{k}-2\varepsilon)E(Q_{j}Q_{k}) - 2\delta_{n} \sum_{k\,:\, k\not =j}
E(Q_{j}Q_{k})
\end{align}
except on an event of probability at most $2m \exp\left\{  -2 n \varepsilon
^{2} \right\}  $.
\end{corollary}

\begin{proof}[Proof of Corollary~\ref{corollary:voteBounds}]
This is an immediate consequence of the inequalities
\eqref{eq:upperv-b}-\eqref{eq:lowerv-b} and Hoeffding's inequality, which
ensures that, for each alternative $j\in [m]$,
\begin{displaymath}
P \xset {
\left|
\sum_{i=1}^{n} U^{i}_{j} -n \mu_{j}
\right|\geq n \varepsilon}\leq 2 e^{-2n \varepsilon^{2}}.
\end{displaymath}
\end{proof}

\begin{proof} [Proof of Corollary~\ref{corollary:BetterVBounds}]
Since $\varepsilon\leq \Delta/4$, and since the means $\mu_{j}$ are
strictly decreasing, we have
\begin{align*}
\mu_{j}- \mu_{k} - 2\varepsilon &\geq \Delta /2 \quad \textrm {for
all} \;\; j<k  \quad \textrm {and}\\
\mu_{j}-\mu_{k} +2 \varepsilon &\leq-\Delta/2  \quad \textrm {for
all} \;\; j>k ,
\end{align*}
or equivalently
\begin{align*}
1 & \leq +2 \Delta^{-1} (\mu_{j}- \mu_{k} - 2\varepsilon) \quad \textrm {for
all} \;\; j<k  \quad \textrm
{and}\\
1 &\leq -  2 \Delta^{-1} (\mu_{j}-\mu_{k} +2 \varepsilon ) \quad \textrm {for
all} \;\; j>k .
\end{align*}
Multiplying through by $2\delta_{n}E(Q_{j}Q_{k})$ shows that
\begin{align*}
2\delta_{n}E(Q_{j}Q_{k}) &\leq +4 \delta_{n}\Delta^{-1} (\mu_{j}- \mu_{k} -
2\varepsilon)  E(Q_{j}Q_{k})  \quad \textrm {for
all} \;\; j<k  \quad \textrm
{and}\\
2 \delta_{n}E(Q_{j}Q_{k})&\leq - 4 \delta_{n}\Delta^{-1} (\mu_{j}- \mu_{k} +
2\varepsilon) E(Q_{j}Q_{k}) \quad \textrm {for
all} \;\; j>k .
\end{align*}
Substituting these upper bounds for the terms
$ 2 \delta_{n}E(Q_{j}Q_{k})$ in the second sum on the the right side
of \eqref{eq:v-upper} indexed by $k$ now shows that the right side
of \eqref{eq:v-upper}  is majorized by
\begin{align*}
&(1+ \delta_{n}-4 \delta_{n}\Delta^{-1})\sum_{k=1}^{j-1}(\mu_{j}-\mu_{k}+
2\varepsilon )E(Q_{j}Q_{k})  \\
+&(1+ \delta_{n}+4 \delta_{n}\Delta^{-1})\sum_{k=j+1}^{m}(\mu_{j}-\mu_{k}+
2\varepsilon )E(Q_{j}Q_{k})\\
\leq &(1- 3 \Delta^{-1}\delta_{n})\sum_{k=1}^{j-1}(\mu_{j}-\mu_{k}+
2\varepsilon )E(Q_{j}Q_{k})  \\
+&(1+ 5 \Delta^{-1}\delta_{n})\sum_{k=j+1}^{m}(\mu_{j}-\mu_{k}+
2\varepsilon )E(Q_{j}Q_{k})  .
\end{align*}
(Here we have used the fact that $0< \Delta \leq 1$, which implies
that $1 \leq \Delta^{-1}$.)  Thus, the upper bound
\eqref{eq:v-upper-improved} follows from \eqref{eq:v-upper}. A
similar calculation shows that the lower bound in
\eqref{eq:v-lower-improved} does not exceed the lower bound in
\eqref{eq:v-lower}. Thus, Corollary~\ref{corollary:BetterVBounds}
follows from Corollary~\ref{corollary:voteBounds}.
\end{proof}

\begin{proof} [Proof of Corollary~\ref{corollary:extremes}]
The hypothesis on $\varepsilon$ ensures that $\mu_{j}-\mu_{j+1}-2
\varepsilon> \Delta/2>0$ for every $j \leq m-1$. Since
$\delta_{n}\downarrow 0$, it follows that for all sufficiently large
$n$ the right side of inequality \eqref{eq:v-lower} for the index
$j=1$ will be \emph{positive}, and similarly the right side of
\eqref{eq:v-upper} for $j=m$ will be \emph{negative}. Hence, for
such $n$ the vote totals $V_{1}$ and $V_{m}$ will satisfy
$V_{m}<0<V_{1}$ on the event that the inequalities
\eqref{eq:v-upper}--\eqref{eq:v-lower} hold.
\end{proof}

\begin{proof} [Proof of Corollary~\ref{corollary:bigQQ}]
Suppose  that there is no sequence
$\theta_{n}\rightarrow \infty$ for which \eqref{eq:bigQQ} is true.
Then for some $\theta<\infty$ there exists a sequence
$(n_{\lambda})_{\lambda\in \mathbb{N}}$ of integers satisfying
$n_{\lambda}\uparrow\infty$ and BN equilibria for these sample sizes
such that
\begin{displaymath}
n_{\lambda} \max_{k\not= j}E(Q_{j}Q_{k})\leq \theta \quad \textrm {for all} \;
\lambda \in \mathbb{N}.
\end{displaymath}
By Corollary~\ref{corollary:BetterVBounds}, it  then follows
that the absolute vote totals $|V_{j}|$ are with high probability
uniformly bounded along the sequence $n_{\lambda}$, that is, for
some $\kappa< \infty$ independent of $\lambda$,
\begin{displaymath}
|V_{j}| \leq \kappa \quad \textrm {for all} \; j \in [m] \textrm
{and all} \lambda\in\mathbb{N},
\end{displaymath}
except on an event of probability at most $2m \exp \xset {-2\varepsilon^{2}n_{l}}$.
But since the outcome probabilities $Q_{j}$ are proportional to the
exponentiated vote totals $e^{V_{j}}$, this  implies that for
every pair $k\not=j$,
\begin{displaymath}
\frac{Q_{k}}{Q_{j}}\leq e^{2 \kappa}
\end{displaymath}
except on an event of vanishingly small probability. This
contradicts Corollary~\ref{corollary:likely-winner}.
\end{proof}

The proof of Proposition \ref{proposition:bigQ1} relies on the following lemma.

\begin{lemma}
\label{lemma:V0} For any constant $\kappa< \infty$, if in some BN equilibrium
for some sample size $n$ there is an alternative $j\in[m]$ such that
\begin{equation}
\label{eq:lemmaV0}\max_{k\,:\, k>j} nE(Q_{j}Q_{k})\leq\kappa,
\end{equation}
then
\begin{gather}
\label{eq:claim-0}P \left\{  V_{j}\geq b \kappa\right\}  \leq2m \exp\left\{
-\Delta^{2}n/32 \right\}  \quad\text{where}\\
b:= (m-1)\sup_{n^{\prime}\geq1}(1+5\Delta^{-1}\delta_{n^{\prime}}%
)(1+\Delta/4)/(2c).\nonumber
\end{gather}

\end{lemma}

\begin{proof}
This is an easy consequence of the upper bound
\eqref{eq:v-upper-improved} of
Corollary~\ref{corollary:BetterVBounds}, using $\varepsilon
=\Delta/8$. Together with the hypothesis \eqref{eq:lemmaV0}, this
implies (taking only the positive terms $k>j$ in
the bound \eqref{eq:v-upper-improved}) that
\begin{align*}
2cV_{j}&\leq n(1+5 \Delta^{-1}\delta_{n})
\sum_{k=j+1}^{m}(\mu_{j}-\mu_{k}+\Delta/4)E(Q_{j}Q_{k})\\
&\leq (1+5\Delta^{-1}\delta_{n})(m-1) (1+\Delta/4)\kappa \\
&\leq  \kappa (m-1)  (1+\Delta/4)\sup_{n' \in\mathbb{N}}(1+5\Delta^{-1}\delta_{n'})
\end{align*}
except on an event of probability at most
$2m \exp \xset {-\Delta^{2}n/32}$.  (Note that because
$\delta_{n}\downarrow 0$, the supremum is
finite.)
\end{proof}

\begin{proof}
[Proof of Proposition~\ref{proposition:bigQ1}] Suppose to the
contrary that for some constant $\kappa<\infty$ and some sequence
$(n_{\ell })_{\ell \in \mathbb{N}}$ of sample sizes there are BN
equilibria such that
\begin{equation}
\label{eq:bad-hypothesis}
\max_{ j \geq 2}n_{\ell }E(Q_{1}Q_{j})\leq \kappa.
\end{equation}
We will show that this hypothesis leads to a contradiction.
Assuming \eqref{eq:bad-hypothesis}, we will prove by induction on
$k$ that for each $0\leq k\leq m-2$ that there is a nonnegative constant
$\kappa_{k}<\infty$,  depending on $\kappa$ but not on the sample
size  $n_{\ell }$
or the BN equilibrium, such that for every $\ell $,
\begin{equation}
\label{eq:induction}
P \xset {\max _{m-k \leq j  \leq m}V_{j}\geq \kappa_{k}} \leq 2m(k+1) \exp
\xset {-\Delta^{2}n_{\ell } /32} .
\end{equation}
The case $k=0$ follows directly from Corollary
\ref{corollary:extremes}, with $\kappa_{0}=0$ and
$\varepsilon=\Delta/8$. Assume, then, that \eqref{eq:induction} is
true for some $k \leq m-2$; we will show that there exists
$\kappa_{k+1}<\infty$ not depending on $n$ or the particular BN
equilibrium such that \eqref{eq:induction} holds when $k$ is
replaced by $k+1$.
For any $\kappa_{*}<\infty$, the ratio $Q_{j}/Q_{1}$ is bounded by
$e^{\kappa_{*}}$ when $V_{1}\geq 0$ and $V_{j}\leq
\kappa_{*}$. Consequently, for any $j'$, with $B= \xset {V_{1}<0}$
and $B_{*}=\xset {V_{j}> \kappa_{^{}}}$,
\begin{align}
\notag
E(Q_{j}Q_{j'})&=
E(Q_{j}Q_{j'}\mathbf{1}_{B^{c}}\mathbf{1}_{B_{*}^{c}})+E(Q_{j}Q_{j'}\mathbf{1}_{B\cup
B_{*}})
\\
\notag       &\leq
E(Q_{j}Q_{j'}\mathbf{1}_{B^{c}}\mathbf{1}_{B_{*}^{c}})+P(B)+P(B_{*})
\\
\label{eq:E-bound}
&\leq e^{\kappa_{*}}E(Q_{1}Q_{j'})+ P \xset {V_{1}<0 } +
P \xset { V_{j}> \kappa_{*}}.
\end{align}
It  follows, by the induction hypothesis ~\eqref{eq:induction} and
Corollary~\ref{corollary:extremes}, that for  every $j \geq m-k$,
\begin{align*}
n_{\ell }E(Q_{m-k-1}Q_{j})&\leq
n_{\ell }e^{\kappa_{k}}E(Q_{1}Q_{m-k-1})
+2 n_{\ell } m(k+2)\exp \xset {- \Delta^{2}n_{\ell }/32}  \\
&\leq \kappa e^{\kappa_{k}} + 2m(k+2)\sup_{n \in
\mathbb{N}}(n\exp \xset {- \Delta^{2}n/32}) \\
&:= C_{k}.
\end{align*}
Observe that the constant $C_{k}$ does not depend on $n_{\ell }$ or
on the particular BN equilibrium.
Lemma~\ref{lemma:V0} now implies that
\begin{displaymath}
P \xset {V_{m-k-1}\geq bC_{k}}\leq 2m \exp \xset
{-\Delta^{2}n_{\ell }/32}.
\end{displaymath}
Set $\kappa_{k+1}= \max (\kappa_{k}, bC_{k})$; then by
\eqref{eq:induction},
\begin{align*}
P \xset {\max _{m-k-1\leq j \leq m} V_{j}\geq \kappa_{k+1}}
&\leq P \xset {\max _{m-k\leq j \leq m }V_{j}\geq \kappa_{k}}+ P
\xset {V_{m-k-1}\geq bC_{k} } \\
& \leq 2m (k+1) \exp \xset {-\Delta^{2}n/32}+2m  \exp \xset {-\Delta^{2}n/32}.
\end{align*}
This completes the induction, and therefore proves that
\eqref{eq:induction} holds for all $k \leq m-2$.
Together with Corollary~\ref{corollary:extremes}, the inequality
\eqref{eq:induction}, with $k=m-2$, implies that
\begin{equation}
\label{eq:Vmax}
P \xset {V_{1}<0 } +P \xset { \max_{j \geq 2}V_{j}\geq
\kappa_{m-2}}\leq 2m^{2}\exp  \xset {-\Delta^{2}n_{\ell }/32}.
\end{equation}
Hence, by  \eqref{eq:E-bound}, for every alternative $j \geq 2$
and $k\not=j$,
\begin{displaymath}
n_{\ell }E(Q_{j}Q_{k})\leq n_{\ell }e^{\kappa_{m-2}}E(Q_{1}Q_{k})
+2m^{2} n_{\ell }\exp \xset {-\Delta^{2}n_{\ell }/32} .
\end{displaymath}
The hypothesis \eqref{eq:bad-hypothesis} therefore implies that for
every $\ell $
\begin{displaymath}
n_{\ell }\max_{j \geq 1}\max_{k\not =j}E(Q_{j}Q_{k})\leq
e^{\kappa_{m-2}}\kappa + 2m^{2}\sup_{n \in \mathbb{N}}(n \exp
\xset {-\Delta^{2}n/32}) .
\end{displaymath}
This contradicts Corollary~\ref{corollary:bigQQ}. Thus, the
hypothesis \eqref{eq:bad-hypothesis} is untenable.
\end{proof}

\begin{proof} [Proof of Proposition~\ref{proposition:Vsmall}]
Once again we will proceed by induction, making use of the result of
Proposition~\ref{proposition:bigQ1}, to prove that for
each $0\leq k \leq m-2$,
\begin{equation}
\label{eq:induction-c2}
P \xset {\max _{m-k\leq j\leq m}V_{j}>1}\leq2m(k+1) \exp \xset
{-\Delta^{2}n/32}
\end{equation}
provided $n$ is sufficiently large.
The case $k=0$ is true by Corollary \ref{corollary:extremes}, with
$\varepsilon=\Delta/8$. Assume, then, that \eqref{eq:induction-c2}
holds for some $k \leq m-3$. We will prove that it is then also true
for $k+1$.
First, observe that without loss of generality we may assume (for
convenience) that all of the sample sizes $n$ are sufficiently large
that $(1-3\Delta^{-1}\delta_{n})\geq 1/2$ and
$(1+5 \Delta^{-1}\delta_{n})\leq 2$. Now because all of the terms in
the lower bound \eqref{eq:v-lower-improved} for $V_{1}$ are
positive, Corollary~\ref{corollary:BetterVBounds} implies that
\begin{align*}
2cV_{1}&\geq n(1-3 \Delta^{-1}\delta_{n}) \sum_{j\geq
2}(\mu_{1}- \mu_{j}- \Delta/8)E(Q_{1}Q_{j}) \\
& \geq n(1-3 \Delta^{-1}\delta_{n}) (\Delta/2) \max _{j
\geq 2}E(Q_{1}Q_{j})\\
& \geq (\Delta/4) \xi_{n},
\end{align*}
where $\xi_{n}$ are the constants in Proposition~\ref{proposition:bigQ1},
except on an event $B$ of probability $P(B)\leq 2m \exp \xset
{-\Delta^{2}n/32}$.
Next, denote by $B_{k}$ the event that $\max
_{m-k\leq j\leq m}V_{j}>1$. By the induction hypothesis,
$P(B) +P(B_{k})\leq 2m(k+2) \exp \xset {-\Delta^{2}n/32}$.
On the complementary event $B^{c}\cap B_{k}^{c}$, we have
\begin{equation}
\label{eq:V-V}
V_{1}-V_{j}\geq \frac{\Delta \xi_{n }}{8c} -1:=
\kappa_{n } \quad \Longrightarrow \quad Q_{j}\leq e^{-\kappa_{n}}Q_{1}
\end{equation}
for all $j \geq m-k$.
Consequently,
\begin{align}
\notag
E(Q_{m-k-1}Q_{j})&=
E(Q_{m-k-1}Q_{j}\mathbf{1}_{B^{c}\cap B_{k}^{c}})+
E(Q_{m-k-1}Q_{j}\mathbf{1}_{B\cup B_{k}})\\
\label{eq:EQQ}
&\leq e^{-\kappa_{n }}E(Q_{1}Q_{m-k-1})
+ 2m(k+2)e^{-\Delta^{2}n/32}
\end{align}
for all $j \geq m-k$. Now the upper bound
\eqref{eq:v-upper-improved} of
Corollary~\ref{corollary:BetterVBounds} (with
$\varepsilon = \Delta/8$) implies that
\begin{align}
\notag
2cV_{m-k-1}&\leq n (1-3\Delta^{-1}\delta_{n_{\ell
}})(\mu_{m-k-1}-\mu_{1}+\Delta/4)E(Q_{1}Q_{m-k-1}) \\
\label{eq:lastVbound}
&+ n(1+5 \Delta^{-1}\delta_{n})
\sum_{j=m-k}^{m}(\mu_{m-k-1}-\mu_{j}+\Delta/4)E(Q_{j}Q_{m-k-1}) .
\end{align}
(Note: This is where the induction would break down if $k=m-2$,
because in this case $m-k-1=1$ and so the upper bound
\eqref{eq:v-upper-improved} would not include the first term.)
Substituting the upper bounds  \eqref{eq:EQQ} for the factors
$E(Q_{j}Q_{m-k-1})$ in  \eqref{eq:lastVbound}, and using the
inequalities
$\mu_{j}-\mu_{1}+\Delta/4 \leq -3\Delta/4$ and
$\mu_{m-k-1}-\mu_{j}+\Delta /4 \leq 5\Delta/4$, we obtain
\begin{align}
\notag
2cV_{m-k-1}\leq &-n (1/2) (3\Delta /4) E(Q_{1}Q_{m-k-1}) \\
\notag
&+ n  (2 (k+1))(5\Delta /4)
e^{-\kappa_{n}}E(Q_{1}Q_{m-k-1}) \\
\label{eq:VinductUpper}
& + n  (2 (k+1))(5\Delta /4)(2m (k+2)e^{-\Delta^{2}n/32}) ,
\end{align}
except on an event $F_{k}$ of probability  $P(F_{k})\leq 2m \exp
\xset {-\Delta^{2}n/32}$. Since $\kappa_{n
}\rightarrow\infty$, the sum of the first two terms on the right
side of the inequality \eqref{eq:VinductUpper} is negative for all
large $n $, and the third term is eventually less than
$2c$. Thus, for all sufficiently large $n $, we have
\begin{displaymath}
V_{m-k-1}<1
\end{displaymath}
except on the event $F_{k}$, and so
\eqref{eq:induction-c2},
\begin{displaymath}
P \xset {\max_{m-k-1\leq j \leq m}V_{j}>1}\leq
P(F_{k})+P(B_{k})\leq 2m(k+2)\exp \xset {-\Delta^{2} n/32}.
\end{displaymath}
This completes the induction, and therefore proves the proposition.
\end{proof}


\begin{thebibliography}{99}                                                                                               %


\bibitem {}Aumman, Robert. \textquotedblleft Mixed and behavior strategies in
infinite extensive games.\textquotedblright\ \textit{Advances in Game Theory},
Princeton Univ. Press (1964): 627--650.

\bibitem {}Arrow, Kenneth J. \textquotedblleft The property rights doctrine
and demand revelation under incomplete information.\textquotedblright%
\ \textit{Economics and Human Welfare} (1979): 23--39.

\bibitem {} Bergemann, Dirk, and Stephen Morris. \textquotedblleft Robust mechanism design.\textquotedblright \textit{Econometrica} 73.6 (2005): 1771--1813.

\bibitem {} Bognar, Katalin, Tilman Börgers, and Moritz Meyer-ter-Vehn \textquotedblleft An optimal voting procedure when voting is costly.\textquotedblright \textit{Journal of Economic Theory} 159 (2015): 1056--1073.

\bibitem {}B{\"{o}}rgers, Tilman. \textquotedblleft Costly
voting.\textquotedblright\ \textit{American Economic Review} 94.1 (2004): 57--66.

\bibitem {}B\"{o}rgers, Tilman. \textit{An Introduction to the Theory of Mechanism Design.} Oxford University Press, USA, (2015).

\bibitem {}Bouton, Laurent. \textquotedblleft A theory of strategic voting in
runoff elections.\textquotedblright\ \textit{American Economic Review} 103.4
(2013): 1248--88.

\bibitem {}Brandt, Felix, Vincent Conitzer, Ulle Endriss, J\'{e}r\^{o}me Lang,
and Ariel D. Procaccia, eds. 2016. \textit{Handbook of Computational Social
Choice}. Cambridge University Press.

\bibitem {}Br\^{a}nzei, Simina, Ioannis Caragiannis, Jamie Morgenstern, and
Ariel D. Procaccia. 2013. \textquotedblleft How bad is selfish
voting?\textquotedblright\ \textit{Twenty-Seventh AAAI Conference on
Artificial Intelligence}.

\bibitem {}Casella, Alessandra. \textquotedblleft Storable
votes.\textquotedblright\ \textit{Games and Economic Behavior} 51.2 (2005): 391--419.

\bibitem {}Casella, Alessandra, Aniol Llorente-Saguer, and Thomas R. Palfrey. \textquotedblleft Competitive equilibrium in markets for votes." \textit{Journal of Political Economy} 120. 4 (2012): 593--658.

\bibitem {}Casella Alessandra, and Antonin Mace. \textquotedblleft Does vote
trading improve welfare?\textquotedblright\ \textit{Annual Review of
Economics} 13 (2021): 57--86.

\bibitem {}Casella, Alessandra, and Luis S{\'{a}}nchez. \textquotedblleft
Storable votes and quadratic voting. An experiment on four California
propositions.\textquotedblright\ {Journal of Politics} 84.1 (2022): 607--612.

\bibitem {}Clarke, Edward H. \textquotedblleft Multipart pricing of public
goods.\textquotedblright\ \textit{Public Choice} 11.1 (1971): 17--33.

\bibitem {}Conitzer, Vincent, Tuomas Sandholm, and J\'{e}r\^{o}me Lang.
\textquotedblleft When are elections with few candidates hard to
manipulate?\textquotedblright\ \textit{Journal of the ACM} 54.3 (2007): 14.

\bibitem {}Dahl, Robert Alan. \textit{Democracy and its Critics.} Yale
University Press (1989).

\bibitem {}d'Aspremont, Claude, and Louis-Andre Gerard-Varet.
\textquotedblleft Incentives and incomplete information.\textquotedblright%
\ \textit{Journal of Public Economics} 11.1 (1979): 25--45.

\bibitem {}Drexl, Moritz, and Andreas Kleiner.
\textquotedblleft Why voting? A welfare analysis.\textquotedblright%
\ \textit{American Economic Journal: Microeconomics} 10.3 (2018): 253--271.

\bibitem {}Duverger, Maurice. \textit{Political Parties: Their Organization
and Activity in the Modern State.} London: Metbuen (1951).

\bibitem {}Eguia, Jon X., and Dimitrios Xefteris. \textquotedblleft
Implementation by vote-buying mechanisms.\textquotedblright\  \textit{American Economic Review} 111.9 (2021): 2811-28.

\bibitem {}Faliszewski, Piotr, and Ariel D. Procaccia. \textquotedblleft AI's
war on manipulation: Are we winning?\textquotedblright\ \textit{AI Magazine}
31.4 (2010): 53--64.

\bibitem {}Feller, William. \textit{An Introduction to Probability Theory and
its Applications. Vol. II,} Second edition (1971). John Wiley \& Sons, Inc.,
New York-London-Sydney.

\bibitem {}Glicksberg, Irving L.. \textquotedblleft A further generalization
of the Kakutani fixed point theorem, with application to Nash equilibrium
points.\textquotedblright\ \textit{Proceedings of the American Mathematical
Society}, 3.1 (1952), 170--174.

\bibitem {}Goeree, Jacob K., and Jingjing Zhang. \textquotedblleft One man,
one bid.\textquotedblright\ \textit{Games and Economic Behavior} 101 (2017): 151--171.

\bibitem {}Goeree, Jacob K., Philippos Louis, and Jingjing Zhang.
\textquotedblleft Improving on simple majority voting by alternative voting
mechanisms.\textquotedblright\ \textit{Oxford Research Encyclopedia of
Economics and Finance} (2020).

\bibitem {}Green, Jerry, and Jean-Jacques Laffont. \textquotedblleft On the
revelation of preferences for public goods.\textquotedblright\ \textit{Journal
of Public Economics} 8.1 (1977): 79--93.

\bibitem {}Groves, Theodore. \textquotedblleft Incentives in
teams.\textquotedblright\ \textit{Econometrica} 41.4 (1973): 617--631.

\bibitem {}Grüner, Hans Peter, and Thomas Tröger. \textquotedblleft Linear voting rules." \textit{Econometrica} 87.6 (2019): 2037--2077.

\bibitem {}Harsanyi, John C. \textquotedblleft Games with incomplete information played by “Bayesian” players, Part I. The basic model."\textit{Management Science} 14.3 (1967): 159--182.

\bibitem {}Harsanyi, John C. \textquotedblleft Games with incomplete information played by “Bayesian” players part II. Bayesian equilibrium points." \textit{Management Science} 14.5 (1968): 320--334.

\bibitem {}Herrera, Helios, Massimo Morelli, and Thomas Palfrey. \textquotedblleft Turnout and power sharing." \textit{The Economic Journal} 124.574 (2014): F131--F162.

\bibitem {}Hoeffding, Wassily. \textquotedblleft Probability inequalities for
sums of bounded random variables.\textquotedblright\textit{Journal of the
American Statistical Association} 58 (1963): 13--30.

\bibitem {}Hylland, Aanund, and Richard Zeckhauser. \textquotedblleft A mechanism for selecting public goods when preferences must be elicited.\textquotedblright \textit{Kennedy School of Government Discussion Paper} Dec. 1980.

\bibitem {}Korolev, V. Yu, and Irina G. Shevtsova. \textquotedblleft On
the upper bound for the absolute constant in the Berry-Esseen
inequality.\textquotedblright\ \textit{Theory of Probability and Its
Applications}, 54 (2010): 638--658.

\bibitem {}Kazumura, Tomoya, Debasis Mishra, and Shigehiro Serizawa.
\textquotedblleft Mechanism design without quasilinearity.\textquotedblright%
\ \textit{Theoretical Economics} 15.2 (2020): 511--544.

\bibitem {}Krishna, Vijay, and John Morgan. \textquotedblleft Overcoming
ideological bias in elections.\textquotedblright\ \textit{Journal of Political
Economy} 119.2 (2011): 183--211.

\bibitem {}Krishna, Vijay, and John Morgan. \textquotedblleft Majority rule
and utilitarian welfare.\textquotedblright\ \textit{American Economic Journal:
Microeconomics} 7.4 (2015): 339--75.

\bibitem {}Lalley, Steven P., and E. Glen Weyl. \textquotedblleft Quadratic
voting: How mechanism design can radicalize democracy.\textquotedblright%
\ \textit{AEA Papers and Proceedings} 108 (2018):\ 33--37.

\bibitem {}Lalley, Steven, and E. Glen Weyl. \textquotedblleft Nash equilibria
for quadratic voting.\textquotedblright\ Available at SSRN 2488763 (2019).

\bibitem {}Ledyard, John O. \textquotedblleft The pure theory of large
two-candidate elections.\textquotedblright \textit{Public Choice} 44.1 (1984):
7--41.

\bibitem {}Ledyard, John O., and Thomas R. Palfrey. \textquotedblleft The approximation of efficient public good mechanisms by simple voting schemes.\textquotedblright \textit{Journal of Public Economics} 83.2 (2002): 153--171.

\bibitem {}Martinelli, César, and Thomas R. Palfrey. \textquotedblleft Communication and information in games of collective decision: A survey of experimental results.\textquotedblright In \textit{Handbook of Experimental Game Theory}. Edward Elgar Publishing, 2020.

\bibitem {}Meir, Reshef. \textquotedblleft Strategic voting.\textquotedblright%
\ \textit{Synthesis Lectures on Artificial Intelligence and Machine Learning}
13.1 (2018): 1--167.

\bibitem{}Mertens, Jean-Francois, and Shmuel Zamir. "Formulation of Bayesian
analysis for games with incomplete information." \textit{International Journal of
Game Theory} 14.1 (1985): 1--29.

\bibitem {}Milgrom, Paul R., \ and Robert J. Weber. \textquotedblleft
Distributional strategies for games with incomplete
information.\textquotedblright\ \textit{Mathematics of Operations Research},
10.4 (1985): 619--632.

\bibitem {}Myerson, Roger B. \textquotedblleft Comparison of scoring rules in
Poisson voting games.\textquotedblright\ \textit{Journal of Economic Theory}
103.1 (2002): 219--251.

\bibitem {}Palfrey, Thomas R. \textquotedblleft A mathematical proof of
Duverger's law.\textquotedblright\ (1989). In: \textit{Ordeshook, P. (Ed.), Models
of Strategic Choice in Politics}. Ann Arbor: University of Michigan Press, 69--91.

\bibitem {}Palfrey, Thomas R., and Howard Rosenthal. "Voter participation and strategic uncertainty." \textit{American Political Science Review} 79.1 (1985): 62--78.

\bibitem {}Penrose, Lionel S. \textquotedblleft The elementary statistics of
majority voting.\textquotedblright\ \textit{Journal of the Royal Statistical
Society }109.1 (1946): 53--57.

\bibitem {} Posner, Eric A., and E. Glen Weyl. \textquotedblleft Voting squared: Quadratic voting in democratic politics." \textit{Vanderbilt Law Review} 68 (2015): 441.

\bibitem {}Reny, Philip J.. \textquotedblleft On the existence of pure and
mixed strategy Nash equilibria in discontinuous games.\textquotedblright%
\ \textit{Econometrica}: 67.5 (1999), 1029--1056.

\bibitem {}Royden, Halsey~L. \textit{Real Analysis}, third edition (1988).
Macmillan Publishing, NY.

\bibitem {}Vickrey, William. \textquotedblleft Counterspeculation, auctions
and competitive sealed tenders.\textquotedblright\ \textit{Journal of
Finance}, 16.1 (1961): 8--37.

\bibitem {}Walzer, Michael. \textit{Spheres of Justice: A Defense of Pluralism
and Equality}\emph{.} Basic Books, 1983.

\bibitem {}Weyl, E. Glen. \textquotedblleft The robustness of quadratic
voting.\textquotedblright\ \textit{Public Choice} 172.1--2 (2017): 75--107.

\bibitem {}Wilson, Robert B. \textquotedblleft Game-theoretic analyses of trading processes.\textquotedblright\ \textit{Advances in Economic Theory, Fifth World Congress of the Econometric Society} (1987): 33--70.
\end{thebibliography}
\end{document}